\documentclass[journal,final,twocolumn]{IEEEtran}
\usepackage[left]{lineno}
\usepackage{comment}
 \usepackage[numbers]{natbib}

\usepackage{amssymb} 
\usepackage{mathtools}

\usepackage[linesnumbered,ruled,lined,commentsnumbered]{algorithm2e}

\usepackage{color} 

\usepackage{tikz}
\usepackage{pgfplots}
\usetikzlibrary{shapes,decorations.pathmorphing}
\usetikzlibrary{arrows}

\usepackage{bchart}

\usepackage{booktabs}

\usepackage{calc}

\usepackage{verbatim}

\usepackage{dsfont} 
\usepackage{grffile} 
\usepackage[shortlabels]{enumitem}

\graphicspath{ {./figures/} } 

\newtheorem{assumption}{Assumption}

\usepackage[textwidth=3cm, textsize=footnotesize]{todonotes}

\newcommand{\todocolor}{magenta}
\tikzstyle{todoboxstyle} = [draw=\todocolor, fill=\todocolor!10, very thick,
    rectangle, rounded corners, inner sep=10pt, inner ysep=10pt]
\tikzstyle{fancytitle} =[fill=\todocolor!10, text=\todocolor]

\newcommand{\1}{{\rm 1}\mskip -4,5mu{\rm l} }

\newcommand{\argmin}{\operatornamewithlimits{argmin}}
\newcommand{\argmax}{\operatornamewithlimits{argmax}}

\newcommand{\abs}[1]{\ensuremath{\lvert #1 \rvert}}

\newcommand{\norm}[1]{\ensuremath{|\!| #1 |\!|}}
\newcommand{\normd}[1]{\ensuremath{\norm{#1}^{\dagger}}}
\newcommand{\normsup}[1]{\ensuremath{\hspace{0.5mm}\!|\!| #1 | \! |_{\infty}}}
\newcommand{\normone}[1]{\ensuremath{\hspace{0.5mm}\!|\!| #1 | \! |_{1}}}
\newcommand{\normtwo}[1]{\ensuremath{\hspace{0.5mm}\!|\!| #1 | \! |_{2}}}

\newcommand{\normt}[1]{\normd{#1}}

\newcommand{\inprod}[2]{\ensuremath{\hspace{0.5mm}\langle #1 , \, #2\rangle}}

\newcommand{\R}{\ensuremath{\mathbb{R}}}
\newcommand{\Rp}{\ensuremath{\R^p}}
\newcommand{\Rn}{\ensuremath{\R^n}}
\newcommand{\Rnp}{\ensuremath{\R^{n\times p}}}
\newcommand{\Rpp}{\ensuremath{\R^{p\times p}}}
\newcommand{\otp}{\ensuremath{\{1,\dots,p\}}}
\newcommand{\numbergroups}{\ensuremath{k}}
\newcommand{\otk}{\ensuremath{\{1,\dots,k\}}}
\newcommand{\otn}{\ensuremath{\{1,\dots,n\}}}

\def\E{\mathbb{E}}

\newcommand{\tuningparameter}{\ensuremath{r}}

\newcommand{\dif}{\ensuremath{b}}
\newcommand{\dift}{\ensuremath{b'}}

\newcommand{\vdiff}{\boldsymbol{\nu}}
\newcommand{\vsub}{\ensuremath{\bar{\boldsymbol{\kappa}}}}
\newcommand{\vsubc}{\ensuremath{\tilde{\boldsymbol{\kappa}}}}

\newcommand{\Ldifdelta}{\boldsymbol{\boldsymbol{\delta}}}
\newcommand{\objective}{\ensuremath{\ell}}

\newcommand{\dualpoint}{\boldsymbol{\nu}}
\newcommand{\cd}{\ensuremath{\tilde\dualpoint}}
\newcommand{\td}{\ensuremath{\bar\dualpoint^\tuningparameter}}

\newcommand{\dgap}[2]{\ensuremath{\Delta(#1 , \, #2)}}

\newcommand{\supp}{\operatorname{supp}}

\newcommand{\vdiffh}{\ensuremath{\Ldifdelta_{\estimatorSbar}}}
\newcommand{\vdiffhc}{\ensuremath{\Ldifdelta_{\estimatorSbari}}}

\newcommand{\zerof}[1]{\ensuremath{\mathbf{0}_{#1}}}

\newcommand{\re}{z} 
\newcommand{\cones}{\ensuremath{\mathcal{C}(\targetset)}}
\newcommand{\conesbar}{\ensuremath{\mathcal{C}(\estimatorSbar)}}

\newcommand{\coneparam}{3}

\newcommand{\thresh}{\ensuremath{\mathcal{T}}}
\newcommand{\nobs}{\ensuremath{n}}

\newcommand{\LI}{{LI}}
\newcommand{\CHICHI}{AV-test}
\newcommand{\lasso}{{lasso}}
\newcommand{\grlasso}{{group lasso}}
\newcommand{\fos}{{FOS}}
\newcommand{\grfos}{{group-FOS}}
\newcommand{\grlassocv}{\ensuremath{\textrm{group-lassoCV}{}_{\texttt{SPAMS}}}}

\newcommand{\lcvs}{\ensuremath{\textrm{lassoCV}{}_{\texttt{SPAMS}}}}
\newcommand{\lcvg}{\ensuremath{\textrm{lassoCV}{}_{\texttt{glmnet}}}}

\newcommand{\loglasso}{{log-lasso}}
\newcommand{\loglcvs}{\ensuremath{\textrm{log-lassoCV}{}_{\texttt{SPAMS}}}}
\newcommand{\logfos}{{log-FOS}}

\newcommand{\group}{{\ensuremath{G}}}
\newcommand{\groupj}{\ensuremath{\group^j}}

\newcommand{\groupsizej}{\ensuremath{p_j}}

\newcommand{\tp}{^\top}

\newcommand{\parameter}{\boldsymbol{\beta}}

\newcommand{\nce}[2]{\expandafter\DeclareRobustCommand\csname #1\endcsname{\ensuremath{#2}}}

\nce{tuningparameterset}{\mathcal R}
\nce{tuningparameter}{r}
\nce{tuningparametero}{\tuningparameter^*}
\nce{tuningparameterc}{\tilde\tuningparameter}
\nce{tuningparameterestimatorfinal}{\hat\tuningparameter}

\nce{parameterset}{\mathcal B}
\nce{parameter}{\boldsymbol{\beta}}
\nce{target}{\parameter^*}
\nce{estimator}{\bar\parameter}
\nce{estimatorfinal}{{\tilde\parameter}^{\tuningparameterestimatorfinal}}

\nce{targetset}{{\mathcal S}}
\nce{targetseti}{\targetset^\complement}

\nce{estimatorset}{\mathcal F_{\tuningparameterset}}
\nce{estimatora}{\estimator^\tuningparameter}
\nce{estimatorahalf}{\estimator^{\tuningparameter/2}}
\nce{estimatorac}{\estimator^{\tuningparameterc}}
\nce{estimatorc}{\tilde\parameter}
\nce{estimatorca}{\estimatorc^{\tuningparameter}}
\nce{estimatorcac}{\estimatorc^{\tuningparameterc}}
\nce{estimatorsetc}{\tilde{\mathcal F}_{\tuningparameterset}}

\nce{targettargetset}{\target_{\targetset}}
\nce{targettargetseti}{\target_{\targetseti}}
\nce{estimatoratargetset}{\estimatora_{\targetset}}
\nce{estimatoratargetseti}{\estimatora_{\targetseti}}

\nce{distance}{d}
\nce{functionstat}{f}
\nce{functioncomp}{g}

\nce{functionstata}{\functionstat(\tuningparameter)}
\nce{functionstatao}{\functionstat(\tuningparametero)}
\nce{functioncompa}{\functioncomp(\tuningparameter)}
\nce{functioncompac}{\functioncomp(\tuningparameterc)}
\nce{functioncompao}{\functioncomp(\tuningparametero)}
\nce{design}{X}
\nce{outcome}{{\mathbf y}}
\nce{observation}{{\mathbf x}}
\nce{noise}{{\mathbf u}}
\nce{estimatorS}{\hat\targetset}
\nce{estimatorSi}{\estimatorS^\complement}
\nce{probasuccess}{\rho}
\nce{functionw}{w}
\nce{matrixw}{W}
\nce{matrixwbar}{\overline{W}}
\nce{matrixwtilde}{\widetilde{W}}
\nce{remainder}{h}

\newcommand{\normtwomat}[1]{\ensuremath{\hspace{0.5mm}\!|\!|\!| #1 | \! | \! |_{2}}}
\newcommand{\normsupmat}[1]{\ensuremath{\hspace{0.5mm}\!|\!|\!| #1 | \! | \! |_{\infty}}}
\newcommand{\normonemat}[1]{\ensuremath{\hspace{0.5mm}\!|\!|\!| #1 | \! | \! |_{1}}}

\newcommand{\matrixR}{R}
\newcommand{\matrixQ}{Q}
\newcommand{\matrixRinv}{R^{-1}}
\newcommand{\matrixQinv}{Q^{-1}}
\newcommand{\normfcn}[1]{\ensuremath{|\!|\!| #1 |\!|\!|}_q}
\newcommand{\matrixI}{\operatorname{I}}
\newcommand{\cminlogreg}{e_{\operatorname{min}}}
\newcommand{\irLog}{a}
\newcommand{\constantlogreg}{c_{\operatorname{log}}}
\newcommand{\gradEslog}{\ensuremath{\tilde{\tau}}}
\newcommand{\gradUpbound}{b}
\newcommand{\relog}{\gamma_{\operatorname{log}}}
\newcommand{\diflog}{b}
\newcommand{\difdelta}{\boldsymbol{\boldsymbol{\delta}}}
\nce{estimatorSbar}{\bar\targetset}
\nce{estimatorSbari}{\estimatorSbar^\complement}

\usepackage{color} 
\usepackage[hidelinks]{hyperref}
\usepackage{enumitem}
\newcommand{\subparagraph}{}

\newcommand{\jl}[1]{\color{blue}{\tiny JL: #1}\color{black}}

\newcommand{\MH}[1]{\color{green}``MH: #1"\color{black}}
\newcommand{\Res}[1]{\color{black}#1\color{black}}

\newcommand{\BlackBox}{\rule{1.5ex}{1.5ex}}  
\newenvironment{proof}{\par\noindent{\bf Proof\ }}{\hfill\BlackBox\\[2mm]}
 
\newtheorem{theorem}{Theorem}
\newtheorem{lemma}[theorem]{Lemma}

\newtheorem{corollary}[theorem]{Corollary}

\usepackage{cancel}

\newif\ifgray
\graytrue
\grayfalse
\IfFileExists{BackgroundColor.tex}{\input{BackgroundColor}}{}
\IfFileExists{../BackgroundColor.tex}{\input{../BackgroundColor}}{}
\pgfplotsset{compat=1.16}
\IEEEoverridecommandlockouts
\IEEEpubid{\makebox[\columnwidth]{978-1-5386-5541-2/18/\$31.00~\copyright2018 IEEE \hfill}
\hspace{\columnsep}\makebox[\columnwidth]{ }}

\ifgray
\pagecolor{black!30!white}
\fi

\ifCLASSINFOpdf
\else
\fi
\begin{document}

\title{Balancing Statistical and\\Computational Precision:
A General Theory and Applications to Sparse Regression}
\author{Mahsa Taheri,
        N\'eh\'emy Lim,
    and~Johannes Lederer
 \thanks{\textit{(Corresponding author: Mahsa Taheri)}}
\thanks{M. Taheri and J. Lederer are with the Department of Mathematics,
       Ruhr-University Bochum,
       44801 Bochum, Germany (e-mail: mahsa.taheri@rub.de; johannes.lederer@rub.de).}
 \thanks{N. Lim is with 
       Business and Decision Benelux,
       1200 Woluwe-Saint-Lambert, Belgium (email: nehemy.lim@businessdecision.be).}

}

\markboth{IEEE TRANSACTIONS ON INFORMATION THEORY}%
{Shell \MakeLowercase{\textit{et al.}}: Bare Demo of IEEEtran.cls for Journals}

\maketitle
\IEEEpubidadjcol
\begin{abstract}
Modern technologies are generating ever-increasing amounts of data.
Making use of these data requires methods that are both statistically sound and computationally efficient.
Typically,
the statistical and computational aspects are treated separately.
In this paper, 
we propose an approach to entangle these two aspects in the context of regularized estimation.
Applying our approach to sparse and group-sparse regression,
we show that it can improve on standard pipelines both statistically and computationally.
\end{abstract}


\IEEEpeerreviewmaketitle

\section{Introduction}\label{intro}

     Contemporary data sets are often large and high-dimensional:
     they contain many parameters and samples, and the number of parameters  rivals or even exceeds the number of samples. 
     On the other hand, 
     one can often assume that the data generating model is sparse or group sparse, that is, that only a small number of parameters or a small number of groups of parameters, respectively, are  relevant.
     A standard objective in the analysis of large and high-dimensional data is the selection of these  relevant parameters or groups of relevant parameters in a computationally feasible and mathematically reliable way. 
     We call these objectives feature selection and group-feature selection, respectively.
    
Standard approaches for these objectives 
such as the \lasso~\citep{lasso} and the \grlasso~\citep{bakin1999,YuanLin06} are equipped with abundant statistical theories~\citep{LedererBook, Buhlmann11,wainwright2019high}. But these theories concern minimizers of objective functions that can be optimized only approximately in practice,
and it is unclear how replacing the true minimizers with computational surrogates impacts the theories.
Moreover, these theories often assume that the tuning parameters take specific values that are not necessarily known in practice.
We are, therefore, interested in interlacing statistical aspects,  computational aspects, and tuning parameter calibration.

In this paper,
we interweave optimization theory and statistical theory via oracle inequalities.
This approach allows us to develop a general algorithm and corresponding statistical theories that account for the optimization aspect and for tuning-parameter calibration.
In effect,
we obtain lasso- and group-lasso-type methods that are faster and more accurate than competing methods and equipped with more comprehensive statistical guarantees.

\subsection{Related Work and Contributions}

One basis of our paper are so-called oracle inequalities.
Oracle inequalities are bounds for the statistical errors of estimators \citep{LedererBook}.
Usually, these bounds hold ``with high-probability,''
that is, they hold on an event that is increasing with the sample size.
Two main problems with standard oracle inequalities are that they (i)~assume tuning parameters that depend on unknown parameters (such as the distribution of the noise) and (ii)~do not account for any optimization aspects.

A solution for the first problem has been proposed in~\citet{Chichignoud_Lederer_Wainwright14}.
Indeed, the AV method in 
\citet[Algorithm~1]{Chichignoud_Lederer_Wainwright14} provides tuning-parameter calibration that leads to---in some sense optimal---oracle inequalities \citep[Theorem~3]{Chichignoud_Lederer_Wainwright14}.
The original AV method
was designed for lasso in linear regression; in the meantime, extensions to logistic regression, graphical models, personalized medicine, and refitted estimation in
linear regression have been established \citep{chetelat2017optimal,laszkiewicz2021thresholded,huang2021tuning,Li2019}.
 However,
these theories still assume that the estimators are computed perfectly, that is, they still do not solve the second problem mentioned above.

In Section~\ref{main proposal}, we extend the above works to solve both mentioned problems. 
In other words, we establish a general algorithm (Algorithm~\ref{mainalgorithm}) that satisfies oracle inequalities (Theorem~\ref{mainresult}) accounting both for tuning-parameter calibration and for potentially inaccurate optimization.

In Section~\ref{examples},
we exemplify our general approach in linear regression and logistic regression.
We find that the approach not only leads to favorable oracle inequalities
but also makes lasso- and group-lasso-type estimation and feature selection faster and more accurate.

\subsection{Notation}

For any set~$\mathcal A$ and any vector~$\boldsymbol{\beta}\in\Rp$, we denote by~$\boldsymbol{\beta}_{\mathcal A}\in\Rp$ the vector that equals~$\boldsymbol{\beta}$ on the coordinates in~$\mathcal A$ and is zero on the coordinates in the complement of~$\mathcal A$, and for any given matrix~$\design\in\Rnp$, we denote by $\design_{\mathcal A}\in\mathbb{R}^{n\times|\mathcal A|}$ the submatrix of $\design$ with column indices in $\mathcal A$.   
For any vector~$\parameter\in\Rp$, we define the support set of~$\parameter$ as~$\supp(\parameter):=\bigl\{j\in\otp:\beta_j\neq 0\bigr\}$. 
We denote the $\ell_q$-norm for vectors by~$\norm{\cdot}_q$ for $q\in[1,\infty]$.
For any matrix~$Q\in\Rnp$, we define the matrix norm induced by the $\ell_q$-norm, denoted by~$\normfcn{\cdot}$ as follows:
\begin{linenomath}
\begin{equation*}
\normfcn{Q}:= \sup_{\observation\neq\zerof{p}}\frac{\norm{Q\observation}_q}{\norm{\observation}_q}\,.
\end{equation*}
\end{linenomath}
The minimal eigenvalue of a square matrix is denoted by~$\Omega_{\min}(\cdot)$. 

To disentangle theory and practice,
we use bars~$\bar{}$~to refer to theoretical estimators (for example, one cannot compute an exact lasso solution in finite time)
and tildes~$\tilde{}$~to practical surrogates (for example, one can approximate a lasso solution within any non-zero tolerance).
To avoid heavy notation, some notation can have different meanings depending on whether we treat linear or logistic regression.

\section{Methodology}\label{main proposal}
 In this section, we state the statistical framework and introduce our estimation approach along with guarantees for it.

\subsection{Motivation}

Consider a standard linear regression model
\begin{linenomath}
\begin{equation*}
  \outcome=\design\target+\noise
\end{equation*}
\end{linenomath}
with $\outcome\in\Rn$, 
$\design\in\Rnp$, 
$\target\in\Rp$,
and~$\noise\in\Rn$.
In genomics,
for example,
\outcome~could contain the levels of a biomarker,
\design~the counts of different SNPs,
\target\ the statistical associations between the SNPs and the biomarker,
and~$\noise$ the measurement errors, batch effects, non-linearities, etc.
The number of SNPs in typical studies is in the millions, 
while the number of study subjects is much smaller: $p\gg n$~\citep{judson2002many}.
We call such settings \emph{high-dimensional}.

In high-dimensional regression, 
classical estimators such as the least-squares estimator are prone to overfitting.
More appropriate are regularized approaches such as the lasso~\citep{lasso}
\begin{linenomath}
\begin{equation*}
  \estimatora\in\argmin_{\parameter\in\Rp}\Bigl\{\frac{1}{2}\normtwo{\outcome-\design\parameter}^2+\tuningparameter\normone{\parameter}\Bigr\}\,,
\end{equation*}
\end{linenomath}
which is a family of $\ell_1$-regularized least-squares estimators indexed by a tuning parameter~$\tuningparameter\in[0,\infty)$.
A feature of the lasso family  is that it can provide small models:
$\#\{(\estimatora)_j\neq 0\}\ll p$.
In genomics, for example, this means that the number of SNPs the estimator deems relevant is small,
which can facilitate interpretation and further processing.
In general, we call the identification of the most relevant predictors \emph{feature selection}.

But regularized approaches such as the lasso spawn three objectives:
\begin{enumerate}
\item Selecting the tuning parameter~\tuningparameter
\end{enumerate}

Large tuning parameters can prevent overfitting, 
but they also introduce a large bias.
A selection scheme must, therefore, balance these two aspects. 

\begin{enumerate}
\setcounter{enumi}{1}
\item Computing the estimators
\end{enumerate}

The lasso's objective function, for example, is convex and ``almost smooth'' but rarely has a minimizer in closed-form.
More generally,
regularized estimators can usually be computed only approximately with descent algorithms such as proximal gradient or coordinate descent.
But these descent algorithms involve another tuning parameter, namely, the number of descent steps.
Picking a large number of steps ensures that the outputs approximate the actual minimizer well,
but this also renders computations slow or even infeasible.
So  a good strategy for calibrating the number of descent steps is needed.

\begin{enumerate}
\setcounter{enumi}{2}
\item Choosing cutoffs for feature selection
\end{enumerate}
Regularized estimators often undergo further treatment such as thresholding.
Thresholding can remove numerical and statistical artifacts and, therefore, reduce the number of false positives.
But thresholding involves yet another tuning parameter, namely, the cutoff.
Large cutoffs remove false positives effectively but can also lead to a large number of false negatives. 
Hence, a sound way to calibrate the cutoff is also in order.

In the context of the lasso,
each of the three objectives individually is understood at least to some extent:
for objective~1,
see \citep{Chichignoud_Lederer_Wainwright14,Bunea},
for example;
for objective~2,
see \citep{Hastie10,osborne2000lasso},
for example;
for objective~3,
see \citep{Chichignoud_Lederer_Wainwright14,Li2019},
for example.
In this paper,
however,
we address all three objectives simultaneously, 
leading to more comprehensive guarantees in theory and to faster and better performances in practice.

\subsection{Statistical Framework}\label{theory}

We now move from the lasso to a more general statistical framework.
Our goal is to estimate an unknown target~$\target\in\parameterset$ in a non-empty set~$\parameterset$.
We consider a family of ``theoretical'' estimators $\estimatorset:=\bigl\{\estimatora:\tuningparameter\in\tuningparameterset\bigr\}$ indexed by a non-empty set of tuning parameters~$\tuningparameterset\subset \R$
and a corresponding family of ``practical'' estimators $\estimatorsetc:=\bigl\{\estimatorca:\tuningparameter\in\tuningparameterset\bigr\}$.
The rationale beyond this distinction is that statistical theory in high dimensions is typically formulated for infeasible estimators  (such as the lasso)
but that these infeasible estimators can be approximated well by practical algorithms (such as proximal gradient descent). 

Our starting point is the statistical and computational theory that is typically available in high dimensions.
\begin{assumption}[Theory basis]\label{feasibility}
Consider a function $\distance:\parameterset\times\parameterset\to[0,\infty)$ that is symmetric and satisfies the triangle inequality,
and consider functions~$\functionstat,\functioncomp:\tuningparameterset\to[0,\infty)$ that are increasing.
Given an oracle parameter~$\target\in\parameterset$ and an oracle tuning parameter~$\tuningparametero\in\tuningparameterset$,
we assume 
the ``statistical'' bound
\begin{linenomath}
 \begin{equation}\label{assumptionstat}
    \distance(\estimatora,\target)\leq \functionstata~~~~~~\forall\, \tuningparameter\geq \tuningparametero
  \end{equation}
  \end{linenomath}
and the ``computational'' bound 
\begin{linenomath}
  \begin{equation}\label{assumptioncomp}
    \distance(\estimatorca,\estimatora)\leq \functioncompa~~~~~~\Res{\forall\, \tuningparameter\geq \tuningparametero}\,.
  \end{equation}
  \end{linenomath}
\end{assumption}
\indent Broadly speaking,
Assumption~\eqref{assumptionstat} ensures that there is a theoretical guarantee for large enough tuning parameters.
Such guarantees are abundant in high-dimensional statistics~\citep{LedererBook}.

Assumption~\eqref{assumptioncomp} ensures that the theoretical estimators can be approximated well.
In other words, Assumption~\eqref{assumptioncomp} specifies what we mean by ``practical estimators:''
these are estimators that are close to their theoretical counterparts in terms of the distance~$\distance$.
Such guarantees, unfortunately, are rare in the literature.
However,
in the applications in Section~\ref{examples},
we relate these guarantees to guarantees in terms of the objective function,
which can then be readily checked by the algorithms.
It would also be very interesting to relate the $\distance$-guarantees to criteria other than the value in objective function,
such as run times of the algorithm, in future work.
In any case, note that different algorithms might lead to different bounds~$\functioncompa$.

The function~$\distance$ is the distance function.
Typical examples in linear regression are $\distance:(\parameter,\target)\mapsto\norm{\parameter-\target}_q$ for a $q\in[1,\infty]$.
The oracle parameter~$\target$ is the ``true'' parameter or any surrogate of it.
A generic example in linear regression is the true regression vector.

\Res{
The oracle tuning parameter~\tuningparametero\ can be any tuning parameter in~$\tuningparameterset$ that makes Assumption~\ref{feasibility} go through. 
This seems quite abstract here in the general setup but will clarify in the specific applications in the later sections.
In linear regression with the lasso,
for example,
it is straightforward to identify $\tuningparametero=2\normsup{\design\tp \noise}$ (or the closest surrogate of it in $\tuningparameterset$) as a natural choice---we will come back to this on Page~\pageref{tuningpage}.
Important here is that $\tuningparametero$ does not need to be known in practice.
}

\subsection{Proposed Method and Theoretical Results}
We now propose a pipeline that deals with our Objectives 1.--3.~stated earlier.
Given a family of theoretical estimators~\estimatorset\ and a corresponding family of practical estimators~\estimatorsetc, 
a distance function~$\distance$, and functions~$\functionstat$ and $\functioncomp$ as specified in Assumption~\ref{feasibility},
our first goal is selecting a practical estimator~$\estimatorca$ that is guaranteed to have a small distance~$\distance(\estimatorca,\target)$.
The best possible guarantee in the sense of Assumption~\ref{feasibility} is
\begin{linenomath}
\begin{equation*}
  \distance(\estimatorc^{\tuningparametero},\target)\leq\functionstatao+\functioncompao.
\end{equation*}
\end{linenomath}
But since neither~$\target$ nor $\tuningparametero$ are known in practice,
we have to make a data-driven selection of~\tuningparameter.

Our proposal for this selection is summarized in Algorithm~\ref{mainalgorithm}.
The general idea is as follows:
Estimators that correspond to tuning parameters that are sufficiently large are ``steady,''
while estimators that correspond to tuning parameters that are too small behave ``erratically'' in high-dimensional settings (think of unregularized estimators in the limit).
The algorithm detects this transition by comparing differences between estimators.

Algorithm~\ref{mainalgorithm} shares its general idea with the AV approach~\citep[Algorithm~1]{Chichignoud_Lederer_Wainwright14} but generalizes that approach in two aspects:
it generalizes it from the lasso to our broad statistical framework,
and it includes the computational discrepancies~$\functioncomp$.
These generalizations, however, make the proofs considerably more difficult.
The main technical challenge is that we can no longer use the theoretical properties of minima of an objective function (KKT conditions, for example).
We overcome this challenge with careful comparisons of the theoretical and practical estimators.

On a technical level,
for the while-loop to make sense and to ensure that the algorithm converges, 
we assume here and in the following that the set of tuning parameters~$\tuningparameterset$ is finite---but, importantly, it can be arbitrarily large.
(We will show later that the specific construction of this set is of minor importance.)

\IncMargin{1em}
\begin{algorithm}[t]\vspace{1mm}
\SetKwInOut{Input}{Inputs}\SetKwInOut{Output}{Outputs}
\Input{\tuningparameterset,\,\distance,\,\functionstat,\,\functioncomp}
\Output{$\tuningparameterestimatorfinal\in\tuningparameterset,\,\estimatorfinal\in\parameterset$}
\BlankLine
Initialize tuning parameter:~$\tuningparameterc\leftarrow\max\{\tuningparameter\in\tuningparameterset\}$\;
Compute $\estimatorcac$\;
\While{$\tuningparameterc\neq \min\{\tuningparameter\in\tuningparameterset\}$ and $\not\exists\,\tuningparameter\in\tuningparameterset\setminus(-\infty,\tuningparameterc]\, :\,  \distance(\estimatorca,\estimatorcac)> \functionstat(\tuningparameter)+\functioncomp(\tuningparameter)+\functionstat(\tuningparameterc)+\functioncomp(\tuningparameterc)$}{  
Update outputs: $\tuningparameterestimatorfinal\leftarrow\tuningparameterc$; $\estimatorfinal\leftarrow\estimatorcac$\;
Go to next tuning parameter: $\tuningparameterc\leftarrow\max\{\tuningparameter\in \tuningparameterset\setminus[\tuningparameterc,\infty)\}$\;
Compute $\estimatorcac$\;
}
\caption{general approach}\label{mainalgorithm}
\end{algorithm}\DecMargin{1em}

We find the following guarantee for Algorithm~\ref{mainalgorithm}.
\begin{theorem}[Guarantee for~\tuningparameterestimatorfinal\ and~\estimatorfinal]\label{mainresult}
Under Assumption~\ref{feasibility}, 
the outputs~$\tuningparameterestimatorfinal$ and~$\estimatorfinal$ of Algorithm~\ref{mainalgorithm} satisfy 
\begin{linenomath}
\begin{equation}\label{r1}
  \tuningparameterestimatorfinal\leq \tuningparametero\,, 
\end{equation}
\end{linenomath}
\begin{linenomath}
 \begin{equation}\label{r2}
    \distance(\estimatorfinal,\target)\leq 3\functionstatao+3\functioncompao\,,
  \end{equation}
  \end{linenomath}
and
\begin{linenomath}
 \begin{equation}\label{r3}
    \E[\distance(\estimatorfinal,\target)]\leq 3\E\functionstatao+3\E\functioncompao\,.
  \end{equation}
  \end{linenomath}
\end{theorem}
\noindent 
The proof of Theorem~\ref{mainresult} is deferred to Appendix~\ref{app:appendixa}.
There, we also state a generalization that allows for deviations from~$\tuningparameterestimatorfinal$ (see Lemma~\ref{pracresult}).

Inequality~\eqref{r1} states that the estimated tuning parameter is bounded by the optimal tuning parameter;
this result will be useful in thresholding small coefficients. 
Inequality~\eqref{r2} states that the selected estimator reaches the best possible bound up to the constant factor~3.
Inequality~\eqref{r3} states the same in expectation.

Note that Inequalities~\eqref{r1} and~\eqref{r2} are entirely ``deterministic:''
they always hold under the stated assumptions.
All probabilistic aspects are summarized in the typically random quantity~$\tuningparametero$---see the following sections with the examples. 
The feature of being ``deterministic'' keeps the bounds very general and allows us to derive Inequality~\eqref{r3} in a rather trivial fashion.

The oracle quantities~\target\ and~\tuningparametero\ are usually unknown.
But, quite strikingly, our method is guaranteed to be essentially optimal even without knowing these quantities.
(Yet, not knowing~\tuningparametero, we cannot derive confidence intervals from the theorem.)
On the other hand,
the functions~\distance, \functionstat, and~\functioncomp\ need to be specified beforehand.
We exemplify the pipeline for common choices of these functions in the following section.
The choice of the set of tuning parameters~$\tuningparameterset$ is usually secondary:
in practice, one can just choose an ad-hoc, ``standard'' grid over the relevant values.
This is not completely obvious but follows from a detailed look at our theories and the typical way the practical estimators are computed:
In our theoretical guarantees, the set~\tuningparameterset\ only appears through~\tuningparametero, that is, 
the finer the grid~\tuningparameterset, 
the better its best representative~$\tuningparametero\in\tuningparameterset$.
In other words, the finer the grid, the better the guarantees.
In practice,
on the other hand,
fine grids also require the computation of many estimators.
But since the estimators are usually computed with warmstarts,
 additional estimators are typically fast to compute.
Moreover,
given the structure of the algorithm,
estimators with $\tuningparameter\leq\tuningparameterestimatorfinal$ do not need to be computed altogether.
We find in the numerical experiments described in the following section that ``standard'' tuning parameter sets such as logarithmically spaced grids with 100 elements give excellent performance.
Also, additional simulations illustrate the limited influence of the specific choice of the tuning parameter set (see Appendix~\ref{app:B2}).
An open question at this point, however, is how to choose~$\tuningparameterset$ optimally.

The above bounds parallel those in~\citep[Theorem~3]{Chichignoud_Lederer_Wainwright14}. 
But besides generalizing the bounds to our framework,
Theorem~\ref{mainresult} contains another, more subtle, improvement:
while~\citet[Theorem~3]{Chichignoud_Lederer_Wainwright14} set the oracle tuning parameter to quantiles of the effective noise~$2\normsup{\design\tp \noise}$, 
our theorem here also allows for setting $\tuningparametero=2\normsup{\design\tp \noise}$ directly.
The latter choice seems more natural and avoids the introduction of quantiles and corresponding levels.
Moreover, it allows us to state bounds in expectation.

The above theories are not only useful for the selection of the tuning parameter~\tuningparameter,
 but they also motivate approaches to our Objectives 2.~(descent steps) and 3.~(cutoff).
More precisely, the proposed algorithm and its theoretical guarantees motivate a holistic approach that covers all three Objectives~1.--3.~simultenously.
Increasing the number of descent steps improves the approximations, that is, decreases $\functioncomp$,
but also adds a computational burden.
Bounds~\eqref{r2} and~\eqref{r3} motivate to calibrate the number of steps such that $\functioncomp=\functionstat$:
this choice~(i) ensures that the practical estimators are equipped with the same statistical bounds as the theoretical estimators up to a factor of two and, as we will see in the example section, 
(ii)~leads to a large increase in computational speed in comparison to pipelines that attempt to descend ``until convergence''. 

The bounds~\eqref{r2} and~\eqref{r3} also motivate a cutoff for feature selection.
Say $\parameterset\subset\Rp$. 
Feature selection, or more generally group-feature selection consists of estimating the support set~$\targetset:=\supp(\target):=\bigl\{j\in\otk:{\target_{\groupj}}\neq \zerof{\groupsizej}\bigr\}$, where~$\group^1,\dots,\group^\numbergroups$ is a partition of $\otp$ and~$\groupsizej$ is the size of group~$\group^j$. 
Typical measures of estimation accuracy~$\distance$ are the metrics  induced by the $\ell_{q,\infty}$-norm with $q\in[1,\infty]$~\citep{hu2017group}:
\begin{linenomath}
	\begin{multline}\label{groupdistance}
	\distance(\parameter,\parameter')\,:=\,\norm{\parameter-\parameter'}_{q,\infty}
	:=\max_{j\in\otk}\Biggl(\sum_{l\in\groupj}\lvert\beta_{l}-\beta'_{l}\rvert^q\Biggr)^{1/q}\\\forall~\parameter,\parameter'\in\Rp\,.
	\end{multline}
\end{linenomath}
We then propose feature selection with the thresholding scheme
\begin{linenomath}
\begin{equation}\label{groupthresholding}
\estimatorS\,:=\,\bigl\{j\in\otk\,:\,\norm{\estimatorfinal_{\groupj}}_q> 3\functionstat(\tuningparameterestimatorfinal)+3\functioncomp(\tuningparameterestimatorfinal) \bigr\}\,,
\end{equation}
\end{linenomath}
where~\tuningparameterestimatorfinal\ and~\estimatorfinal\ are the outputs of Algorithm~\ref{mainalgorithm}. 
And we can find the following guarantee for~$\estimatorS$.

\begin{lemma}[Group-feature selection]\label{groupres}
	Assume the beta-min condition 
	\begin{linenomath}
	\begin{equation*}
	\min_{j\in\targetset}\norm{\target_{\groupj}}_q>  6\functionstatao+6\functioncompao\,.    
	\end{equation*}
	\end{linenomath}
	Then, it holds that
	\begin{linenomath}
	\begin{equation*}
	\estimatorS\,=\,\targetset\,.
	\end{equation*}
	\end{linenomath}
	
\end{lemma}
\noindent 
Hence, as long as the $\ell_q$-norms of the non-zero groups of the true vector are not too small,
one can safely threshold each group of \estimatorfinal\ with the cutoff $3\functionstat(\tuningparameterestimatorfinal)+3\functioncomp(\tuningparameterestimatorfinal)$ without increasing the number of false negatives.

Note that Lemma~\ref{groupres} as well as Theorem~\ref{mainresult} are stochastic: 
the randomness of the data is reflected in the oracle tuning parameter~\tuningparametero.

\section{Examples}\label{examples}
In this section, we exemplify our general approach.
We consider $\parameterset:=\R^p$ and a measure of accuracy~\distance\ that is induced by a norm: 
$\distance(\parameter,\parameter'):=\norm{\parameter-\parameter'}$.
The theoretical estimators are defined by
\begin{linenomath}
\begin{equation*}
  \estimatora\in\argmin_{\parameter\in\R^p}\objective(\parameter,\tuningparameter)
\end{equation*}
\end{linenomath}
for $\tuningparameter\in\tuningparameterset\subset[0,\infty)$ and $\objective(\parameter,\tuningparameter):=L(\parameter)+\tuningparameter\normt{\parameter}$ with a convex and differentiable function $L\,:\,\R^p\to [0,\infty)$.
Specifically, we take $L$ as the least-squares loss (Section~\ref{LR}) and the logistic loss (Section~\ref{logreg}).
Our  conditions on the regularizer--besides it being a norm is:  \label{decomppage}1.~decomposability with respect to the support~$\targetset=\supp(\target)$ of the target~$\target$, that is,  $\normt{\parameter}=\normt{\parameter_\targetset}+\normt{\parameter_{\targetseti}}$ for all~$\parameter\in\Rp$, which holds for many standard regularizers, for example, see~\citep[Section~2.2]{Negahban12} and \citep[Section~3.2]{Wainwright14} and 2.~the regularizer is the dual norm of   $\norm{\cdot}$ (induced norm in the measure of accuracy~\distance).

\subsection{Linear Regression}\label{LR}
Here, we specify our approach to regularized linear regression.
In Section~\ref{lassomethod}, we provide sufficient conditions for Assumption~\eqref{assumptionstat} and Assumption~\eqref{assumptioncomp}  (see Lemma~\ref{linreg} and Lemma~\ref{combound}, respectively), 
and we exemplify the results by considering group-feature selection with the group lasso.
In Section~\ref{lassoalgo}, we specify the implementation for group-feature selection (see Algorithm~\ref{algo}),
and in Section~\ref{lassosim}, we demonstrate the empirical efficiency of our approach both on simulated and real data. 

\subsubsection{Method}\label{lassomethod}

\nce{compatibility}{c}

We can apply our general scheme in Algorithm~\ref{mainalgorithm} to the linear regression model
\begin{linenomath}
\begin{equation*}
  \outcome=\design\target+\noise\,,
\end{equation*}
\end{linenomath}
where~$\outcome\in\Rn$ is the outcome, $\design\in\Rnp$  the design matrix, $\target\in\Rp=:\parameterset$  the regression vector with support~$\targetset=\supp(\target)\subset\otp$, and~$\noise\in\Rn$  the random noise. 
To estimate~$\target$, we consider the objective function
\begin{linenomath}
	\begin{equation}\label{obj}
	\objective(\parameter,\tuningparameter):=L(\parameter)+\tuningparameter\normt{\parameter}\,,
	\end{equation}
\end{linenomath}
where~$\parameter\in\Rp$, $\tuningparameter\in[0,\infty)=:\tuningparameterset$, and $L(\parameter):=\normtwo{\outcome-\design\parameter}^2/2$. 
We then propose to use the estimators~$\estimatorset=\bigl\{\estimatora:\tuningparameter\in\tuningparameterset\bigr\}$ with
\begin{linenomath}
	\begin{equation}\label{lassoobj}
	\estimatora\in\argmin_{\parameter\in\Rp}\objective(\parameter,\tuningparameter)\,.
	\end{equation}
\end{linenomath}

A sufficient condition for satisfying the statistical bound in Assumption~\eqref{assumptionstat} for the estimators in Display~\eqref{lassoobj} is as follows:
\begin{lemma}[\functionstat\ in regularized linear regression]\label{linreg}
Assume there is a constant~$\compatibility\in(0,\infty)$ such that
\begin{linenomath}
\begin{equation*}
    \norm{\vdiff}\leq \frac{\norm{\design\tp \design\vdiff}}{\nobs\compatibility}~~~\forall\,\vdiff\in\Rp:\normt{\vdiff_{\targetseti}}\leq 3\normt{\vdiff_\targetset}\,.
\end{equation*}
\end{linenomath}
 Then,~\estimatorset\ satisfies
 \begin{linenomath}
  \begin{equation*}
     \norm{\target-\estimatora} \leq \functionstata~~~\forall\, \tuningparameter\geq \tuningparametero
 \end{equation*}
 \end{linenomath}
for~$\functionstata:={3\tuningparameter}/({2\nobs\compatibility})$ and~$\tuningparametero:=2\norm{\design\tp\noise}$.
\end{lemma}
\noindent 
The  condition in the upper display of the lemma is a variant of the classical compatibility  condition~\citep{koltch09a,Sara09}. 
If~$\norm{\cdot}\equiv\normsup{\cdot}$, for example, it coincides with the~$\ell_\infty$-restricted eigenvalue condition formulated in~\citep[Equation~(7)]{Chichignoud_Lederer_Wainwright14};
in this sense, Lemma~\ref{linreg} is a generalization of their Lemma~5. 
We defer the proof of Lemma~\ref{linreg} to Appendix~\ref{app:appendixa}.

\Res{
The lemma also clarifies the choice of the oracle tuning parameter~\tuningparametero\ in linear regression.
Recall that in the abstract setup of Section~\ref{main proposal},
the oracle tuning parameter could be any tuning parameter that makes the assumptions go through; 
\label{tuningpage}
here, we identify $\tuningparametero:=2\norm{\design\tp\noise}$ as a valid choice in linear regression.
In the lasso case, where $\norm{\cdot}\equiv\normsup{\cdot}$, the value of the oracle tuning parameter coincides with the usual value in the literature---see, for example,  \citep[Pages~1--2]{lederer2021estimating}, \citet[Page~9]{Lederer2019}, and \citet[Page~112]{LedererBook}.
Thus,
in line with that standard literature,
we can interpret \tuningparametero\ as the ``best'' tuning parameter in the sense that it is the smallest tuning parameter that controls the effective noise \citep[Section~4]{dalalyan2017prediction}.
}

A sufficient condition for satisfying the computational bound in Assumption~\eqref{assumptioncomp}  for estimators in Equation~\eqref{lassoobj} is as follows:
\begin{lemma}[\functioncomp\ in regularized regression] \label{combound}
For given $\gradUpbound\in[0,\infty)$, assume there is a  constant $\re\in(0,\infty)$   such that
\begin{linenomath}
	\begin{equation*}
	L(\parameter+\vdiff)-L(\parameter)-\inprod{\nabla L(\parameter)}{\vdiff}\geq \re\nobs{\normt{\vdiff}}^2 
	\end{equation*}
	\end{linenomath}
	for all $\parameter\in\Rp$ and $\vdiff\in\Rp$ that satisfy $\normt{\vdiff_{\targetseti}}\leq \coneparam\normt{\vdiff_{\targetset}} +4\normt{\estimator^{\tuningparameter}_{\targetseti}} + \dif/\tuningparameter$.
Assume also that $\norm{\cdot}\le \normt{\cdot}$. 
\Res{Then, we have}
\begin{linenomath}
	\begin{equation*}
	\norm{\estimatora-\estimatorc} \leq \functioncompa~~~\forall\, \Res{\tuningparameter\in [\tuningparametero,\infty)}
	\end{equation*}
	\end{linenomath}
		for $\functioncompa:=\sqrt{\gradUpbound/(\re\nobs)}+\tuningparameter/(\re\nobs)$ and every $\estimatorc\in\Rp$ with $\objective(\estimatorc,2\tuningparameter)\leq \objective(\estimator^{\tuningparameter},2\tuningparameter)+\dif$.
\end{lemma}
\noindent 
\Res{Above lemma gives a tool to check whether the ``computational bound" is reached by using the  objective's value:
  once $\objective(\estimatorc,2\tuningparameter)\leq \objective(\estimator^{\tuningparameter},2\tuningparameter)+\dif$ is verified, then, we can make sure that $\norm{\estimatora-\estimatorc} \leq \functioncompa$ is verified as well for  $\functioncompa:=\sqrt{\gradUpbound/(\re\nobs)}+\tuningparameter/(\re\nobs)$.
}
The result is formulated for general convex and differentiable loss functions to make it useful also for the case of logistic regression in the next section.
The condition in the upper display of the lemma is a slightly stronger version of restricted strong convexity~\citep[Pages~291ff]{Hastie2015},~\citep[Pages~277ff]{wainwright2019high} (see Section~\ref{sec:condition} for details).

Now, plugging Lemma~\ref{linreg} and Lemma~\ref{combound} into Theorem~\ref{mainresult} yields the following result.
\begin{corollary}[Estimation in regularized linear regression]\label{EsLR}
 Suppose that the conditions of Lemma~\ref{linreg} are met and $\functioncomp=\functionstat$. 
Then, the outputs of Algorithm~\ref{mainalgorithm} satisfy
\begin{linenomath}
  \begin{equation*}\label{r11}
  \tuningparameterestimatorfinal\leq \tuningparametero\,, 
\end{equation*}
\end{linenomath}
\begin{linenomath}
 \begin{equation*}\label{r22}
    \distance(\estimatorfinal,\target)\leq 9\tuningparametero/(nc)\,,
  \end{equation*}
  \end{linenomath}
and
\begin{linenomath}
 \begin{equation*}\label{r33}
    \E[\distance(\estimatorfinal,\target)]\leq \E[9\tuningparametero/(nc)]\,.
  \end{equation*}
  \end{linenomath}

\end{corollary}
This equips the proposed algorithm  with guarantees for its estimation error in linear regression models.

Besides estimation, our approach can also be applied to feature selection.
Specifically suited base estimators for feature selection are sparsity-inducing methods, such as the \grlasso:
\begin{linenomath}
\begin{equation*}\label{objgroupfos}
  \estimatora\in\argmin_{\parameter\in\Rp}\Bigl\{\frac{1}{2}\normtwo{\outcome-\design\parameter}^2+\tuningparameter\sum_{j=1}^{\numbergroups}\sqrt{\groupsizej}\normtwo{\parameter_{\groupj}}\Bigr\}\,,
\end{equation*}
\end{linenomath}
which corresponds in our framework to  the distance
\begin{linenomath}
\begin{multline*}\label{grouplassodistance}
  \distance(\parameter,\parameter')\,:=\,\max_{j\in\otk}\normtwo{\parameter_{\groupj}-\parameter'_{\groupj}}/\sqrt{\groupsizej}\\
  \forall~\parameter,\parameter'\in\Rp\,.
\end{multline*}
\end{linenomath}
An estimate of the support with our thresholding scheme is
\begin{linenomath}
\begin{equation}\label{grouplassothresholding}
  \estimatorS\,:=\,\bigl\{j\in\otk\,:\,\normtwo{\estimatorfinal_{\groupj}}/\sqrt{\groupsizej}> 9\tuningparameterestimatorfinal/(\nobs\compatibility) \bigr\}\,,
\end{equation}
\end{linenomath}
where~\tuningparameterestimatorfinal\ and~\estimatorfinal\ are the outputs of Algorithm~\ref{mainalgorithm} with~$\functionstat(\tuningparameter):=g(\tuningparameter):=3\tuningparameter/(2\nobs\compatibility)$.
Assuming the above-mentioned compatibility condition and an equally standard beta-min condition~\citep{LedererBook,Buhlmann11}
\begin{linenomath}
\begin{equation*}
  \min_{j\in\targetset}\normtwo{\target_{\groupj}}/\sqrt{\groupsizej}>  \frac{18\tuningparametero}{\nobs\compatibility}\,,
\end{equation*}
\end{linenomath}
we obtain a sharp guarantee for~$\estimatorS$:

  \begin{corollary}[Group-feature selection in regularized linear regression]\label{grouplassores}
Under the stated conditions, it holds that
\begin{linenomath}
\begin{equation*}
  \estimatorS\,=\,\targetset\,.
\end{equation*}
\end{linenomath}
  \end{corollary}
\noindent This guarantee takes both the tuning parameter calibration and the computational tolerances of the implementations into account;
in other words, 
Equation~\eqref{grouplassothresholding} is a theoretically justified thresholding scheme.

\subsubsection{Algorithm}\label{lassoalgo}

  Algorithm~\ref{algo}  is a specification of  our general scheme for group-feature selection in linear regression. 
Recall that bars~$\bar{}$ refer to theoretical estimators,
while tildes~$\tilde{}$ refer to computable surrogates. 
We call the algorithm~\grfos\ (group-{\it F}ast and {\it O}ptimal {\it S}election) for convenient reference. 
The three challenges described earlier are addressed as follows: 
(1)~The stopping point on the tuning parameter path is determined via AV-tests~\citep{Chichignoud_Lederer_Wainwright14}, which contrasts estimators in terms of the distance function (line~19 of Algorithm~\ref{algo}). 
Note that here, Equation~\eqref{groupdistance} is used as the distance function.
 (2)~We use the duality gap to ensure that the computational bound is reached.
  For any regression vector~$\estimatorc$ and any feasible dual variable~$\cd$, the duality gap is~$\dgap{\estimatorc}{\cd}:=\objective(\estimatorc,2\tuningparameter)-D(\cd,2\tuningparameter),$
   where~$D(\cd,2\tuningparameter)$ is the  dual function, 
  and it is  well known~\citep{Borwein2010} that~$\dgap{\estimatorc}{\cd}$ is an upper  bound of~$\objective(\estimatorc,2\tuningparameter)-\objective(\estimator^{2\tuningparameter},2\tuningparameter)$. 
  Since $\objective(\estimator^{2\tuningparameter},2\tuningparameter)\le \objective(\estimatora,2\tuningparameter)$, we conclude  that $\dgap{\estimatorc}{\cd}$ is also an upper bound of $\objective(\estimatorc,2\tuningparameter)-\objective(\estimatora,2\tuningparameter)$.
  This upper bound ensures that the required precision is reached. Importantly, we do not need to solve the dual problem of the \grlasso, but instead, we only require a dual point, which can be found with an explicit expression. 
  We refer to Lemma~\ref{combound} to show the connection between computational bound  and  duality gap and Appendix~\ref{app:B1} regarding the construction of feasible dual points.
  Setting~$\functionstata:=g(\tuningparameter):={3\tuningparameter}/({2\nobs\compatibility})$, which, by Lemma~\ref{combound}, gives the stopping criterion as $\dgap{\estimatorc}{\cd}\le \dif:=\tuningparameter^2(3\re/(2\compatibility)-1)^2/(\re\nobs) $.
   As the optimization algorithm, one could select proximal gradient descent~\citep{Beck09}, coordinate descent~\citep{Hastie10}, or other techniques. 
   We opted for the first one; the corresponding updates in line~14 then read for each group
  \begin{linenomath}
\begin{equation*}
    \estimatorc_{\groupj} \mapsto \thresh_{\tuningparameter\sqrt{\groupsizej}/m}(\estimatorc_{\groupj} - \frac{1}{m}\design_{\groupj}^\top(\design\estimatorc-\outcome))\,,
\end{equation*}
\end{linenomath}
 where~$\thresh$ is the block soft-thresholding operator defined by~$\thresh_t(\boldsymbol{a}_{\groupj}) := \max(1-t/\normtwo{\boldsymbol{a}_{\groupj}},0)\boldsymbol{a}_{\groupj}$ for $j\in\otk$ and where~$m\in(0,\infty)$ is the step size determined by backtracking. (3)~We finally use Equation~\eqref{grouplassothresholding}  for thresholding, which guarantees correct (group) feature selection (see Corollary~\ref{grouplassores}).   

Neglecting all the intricate details, 
our method is simply a roughly computed \grlasso\ estimate with subsequent thresholding of the elements.

\begin{algorithm*}[!ht]\vspace{1mm}
\SetKwInOut{Input}{Inputs}\SetKwInOut{Output}{Outputs}
\Input{$\outcome\in\Rn$; $\design\in \Rnp$; $\tuningparameter_1=\tuningparameter_{\max} > \tuningparameter_2 > \ldots > \tuningparameter_M>0$}
\Output{$\estimatorc^{\tuningparameterestimatorfinal}\in\Rp$; $\estimatorS\subset\otk$}
\BlankLine
\textbf{Initialization :} \texttt{statsCont:=true}; \texttt{statsIt:=}1; $\estimatorc^{\tuningparameter_1}:=\zerof{p}$; $\tuningparameterestimatorfinal:=\tuningparameter_M$\; 
\While{\texttt{statsCont==true} \texttt{AND} $\texttt{\textup{statsIt}}\,<M$}{  
    \texttt{statsIt:=statsIt+1}\;
    \texttt{compCont:=true}\;
    \texttt{betaOld:=}$\estimatorc^{\tuningparameter_{\texttt{\textup{statsIt}}-1}}$\;
    \While{\texttt{compCont==true}}{
        Compute a feasible dual point $\cd^{\tuningparameter_\texttt{\textup{statsIt}}}$ \; 
        Compute the duality gap $\dgap{\estimatorc^{\tuningparameter_\texttt{\textup{statsIt}}}}{\cd^{\tuningparameter_\texttt{\textup{statsIt}}}}$\;
        \eIf{$\dgap{\estimatorc^{\tuningparameter_\texttt{\textup{statsIt}}}}{\cd^{\tuningparameter_\texttt{\textup{statsIt}}}}\leq \,\dif$}{
        $\estimatorc^{\tuningparameter_\texttt{\textup{statsIt}}}$\texttt{:=betaOld}\;
        \texttt{compCont:=false}\;
        }{
        \For{$j=1,\ldots, k$}{
        $\estimatorc_{\groupj}^{\tuningparameter_\texttt{\textup{statsIt}}}\texttt{:=}\thresh_{\tuningparameter_\texttt{\textup{statsIt}}\sqrt{\groupsizej}/L}\left(\texttt{betaOld}_{\groupj} - \design_{\groupj}^\top(\design\cdot\texttt{betaOld}-\outcome)/L \right)$\;
        }
      \texttt{betaOld:=}$\estimatorc^{\tuningparameter_\texttt{\textup{statsIt}}}$\;
        }
    }
    \texttt{statsCont:=}$\displaystyle \prod_{i=1}^{\texttt{\textup{statsIt}}} \mathds{1}\bigl\{ \max_{j\in\otk}\normtwo{\estimatorc_{\groupj}^{\tuningparameter_\texttt{\textup{statsIt}}} - \estimatorc_{\groupj}^{\tuningparameter_i}}/\sqrt{\groupsizej}(\tuningparameter_\texttt{\textup{statsIt}} + \tuningparameter_i) - \big(3/(\nobs\compatibility)\big) \leq 0\bigr\}$\;
}
\If{\texttt{statsCont==false}}{
    $\tuningparameterestimatorfinal:=\tuningparameter_{\texttt{\textup{statsIt}}-1}$\;
    }
$\estimatorS\,:=\,\bigl\{j\in\otk:\normtwo{\estimatorc^{\tuningparameterestimatorfinal}_{\groupj}}/\sqrt{\groupsizej}> 9\tuningparameterestimatorfinal/(\nobs\compatibility) \bigr\}$
\caption{group-feature selection in linear regression}\label{algo}
\end{algorithm*}

As initialization, we choose the all-zeros vector in~$\Rp,$ reflecting our assumption that many groups are inactive. 
Since we are limited to finitely many computations in practice, we consider finite sequences~$r_1=r_{\max}>r_2>\dots>\tuningparameter_M=\tuningparameter_{\min}>0.$ The concrete choice follows the ones used in standard implementations~\citep{Hastie10}: we use a logarithmically spaced grid of size~$M=100,$ set~$\tuningparameter_{\max} := \max_{j\in\otk}\normtwo{\design_{\groupj}^\top \outcome}/\sqrt{\groupsizej}$~to the smallest tuning parameter such that~$\estimatora=\zerof{p}$, and  define~$\tuningparameter_{\min} := \tuningparameter_{\max}/h$ as a fraction of~$\tuningparameter_{\max}$. Standard choices for~$h$ range from~$100$ to~$10\,000.$ 
On a very high level,
assuming bounded group sizes, it holds that $\tuningparameter_{\max}/\tuningparametero\approx \max_{j\in\otk}\normtwo{\design_{\groupj}^\top \outcome}/\max_{j\in\otk}\normtwo{\design_{\groupj}^\top \noise}\approx n\sum_{j=1}^k\sqrt{\groupsizej}\normtwo{\target_{\groupj}}/\sqrt{n}\approx \sqrt{n}.$ 
To ensure that~$\tuningparameter_{\min}< \tuningparametero$ on our data sets, we thus select~$h:=1000.$ Moreover, note that our theoretical results hold for any types of grids (also for continuous ranges of \tuningparameter).
 And because of the warm starts and the early stopping, the computational complexity of \grfos\ depends only very mildly on~\mbox{$M$ and~$\tuningparameter_{\min}.$}
Finally,
$\compatibility=2$ and~$\re=1$ are considered global constants.
We show later that changing these constants has minor impact on our results (see Tables~\ref{Const_impact_LR} and~\ref{C_impact_LR}).
 
 Computationally, the proposed method has two main advantages: First, only a part of the tuning parameter path needs to be computed, more precisely, only the part with large and moderate tuning parameters. Second, only very rough computations are required; in particular, since a large tolerance can be accepted for large tuning parameters, only very small numbers  of optimization cycles (in practice, often zero to five) are required per tuning parameter.

\subsubsection{Empirical performance}\label{numerics}\label{lassosim}

We now demonstrate the computational efficiency and the empirical accuracy of the proposed method. To obtain a comprehensive overview, we consider a variety of experiments for feature selection and group-feature selection, including  biological and financial  applications as well as synthetic data.  \grfos\ can be adapted easily to handle data without structure; we call this  version  \fos\  for convenient reference. We compare \fos\ to the \lasso\ with Cross-Validation (lassoCV), which  is currently the most popular method for feature selection, and with the non-convex approaches SCAD~\citep{Fan_Li01} and MCP~\citep{Zhang10}; 
and we compare \grfos\ with the \grlasso\ with Cross-Validation (group-lassoCV).
We start by synthetic data sets to verify the efficiency of \fos\ and \grfos\ in regressions with moderately large data (up to $10\,000$ samples and parameters). 
Then, we  show the scalability of \fos\ by analyzing a financial data set with more than $150\,000$ parameters.
 (The application to even larger regression data is currently limited by the memory restrictions in  \texttt{MATLAB}\textsuperscript{\textregistered}; a future \texttt{C/Fortran} implementation could remove this limitation). 
We continue by learning a biological network with a neighborhood selection scheme. 
Each of the corresponding regressions comprises only 500 samples and 1000 parameters,
 but since 1000 such regressions are needed, the computational complexity can easily render standard methods infeasible. 
 Finally, we show the efficiency of \grfos\ in classification on biological data.

All computations are conducted with \texttt{MATLAB}\textsuperscript{\textregistered} and are run on an Intel Core(TM) i5-3470 CPU(3.20GHz).   \fos\ and \grfos\ are implemented using the \texttt{SPAMS} package~\citep{BJMO11} coded in \texttt{C++}, which uses the duality gap as the convergence criterion. 
The frequency for checking the duality gap is set to every 1 iteration in the optimization function.  \fos\ is compared with two lassoCV implementations: First, lassoCV is implemented analogously to \fos\ using the \texttt{SPAMS} package and a 10-fold Cross-Validation with warm starts. This implementation, called $\lcvs$ in the following, is the most appropriate one for comparisons with the \fos\ implementation. However, much work has gone into efficient implementations of lassoCV. Therefore, we also use the well-known \texttt{glmnet} package~\citep{Hastie10} and call the corresponding implementation~\lcvg. However, these results must be treated with reservation, because \texttt{glmnet} cannot be calibrated to the same convergence criterion as our implementation. More precisely, the convergence criterion in \texttt{glmnet} needs to be specified in terms of maximum change in the objective, which does not coincide with the criterion in our algorithm and with the convergence of the estimator itself as needed in the theory. One could also argue that comparing \fos\ with \lcvg\ is not fair in any case, because \texttt{glmnet} exploits additional geometric properties of the \lasso\ (such as screening rules). These additional properties could also be used in our scheme, but their implementation is deferred to future work. In any case, we demonstrate that even in its current version, \fos\ can outperform both \lcvs\ and~\lcvg. For SCAD and MCP, we use the \texttt{SparseReg} toolbox~\citep{Zhou12}. The tuning parameters that balance the fitting and penalty are set via Cross-Validation. However, SCAD and MCP each also contain a second tuning parameter that determines the shape of the penalty: for SCAD, it is set to 3.7, which  minimizes the Bayes risk; for MCP, it is set to 1, the default value in the \texttt{SparseReg} toolbox. 
\grlassocv\ is implemented analogously to \grfos, using the \texttt{SPAMS} package and a 10-fold Cross-Validation with warm starts. 

Code can be found under \href{https://github.com/LedererLab}{github.com/LedererLab}.

\paragraph{Synthetic data}\label{synthetic}
In this part, we use synthetic data to demonstrate the empirical performance of \fos\ and \grfos. 
For \fos,
data are generated from linear regression models with~$\nobs=500$ and~$p=1000$ and with~$\nobs=5000$ and~$p=10\,000$.
Each row of the design matrix~$\design\in\Rnp$ is sampled independently from a~$p$-dimensional normal distribution with mean~$0$ and covariance matrix~$(1-\rho)\matrixI+\rho\1,$ where~$\matrixI$ is the identity matrix, $\1$  the matrix of ones, and~$\rho=0.3$  the correlation among the features. 
The design matrix is then normalized such that its columns have Euclidean norm equal to~$\sqrt n.$ 
The entries of the noise~$\noise\in\Rn$ are generated according to a 
one-dimensional standard normal distribution. 
The entries of~$\target$ are first set to~$0$ except for~$10$
uniformly at random chosen entries that are each set to~$1$ or~$-1$
with equal probability;
the whole vector~$\target$ is then rescaled such
that the signal-to-noise ratio~${\|X \target\|_2^2/\nobs}$ is equal to 5.

For \grfos, the same setup is used except for the selection of non-zero parameters: 
the index set is partitioned into groups of equal length,
and the non-zero parameters are those in the  uniformly at random chosen ``active'' groups.
To obtain a wider range of cases,
data size, group lengths, and numbers of activated groups are also modulated. 
\begin{table*}
\caption{Average run times (in seconds), Hamming distances, and Estimation errors for \lcvs, \lcvg, SCAD, MCP, and  \fos, with $M=100$.
 For the larger data set, \lcvs\ timed out on our machine, which means that \lcvs\ took more than one hour for the first run.
The results illustrate that \fos\ is both faster and almost more accurate than  other methods across all settings }
 \label{tab:time_hamming} 
\centering
  \begin{tabular}{ c c c c c  c c}
    \toprule
  & \multicolumn{2}{c}{$~\nobs=500,\,p=1000$} & \multicolumn{4}{c}{$\nobs=5000,\,p=10\,000$}                  \\
	\cline{2-7}
   Method     & Timing     & Hamming distance & Estimation error & Timing     & Hamming distance & Estimation error\\
	\hline
    \lcvs & $132.72\pm 16.14$  &  $71.10\pm18.17$ & $0.19\pm0.04$  &  NA  & NA & NA\\
    \lcvg     & $\phantom{00}3.61\pm \phantom{0}1.10$ & $55.10\pm 18.89$  & $0.20\pm0.04$   & $\phantom{00}32.92\pm \phantom{0}7.12$ & $\phantom{0}56.80\pm 25.42$  & $0.07\pm0.01$ \\
   SCAD & $103.50\pm \phantom{0}4.05$  &  $70.40\pm23.51$ & $0.20\pm0.03$  & $1410.60\pm87.40$  &$145.40\pm43.96$ & $0.08\pm0.00$\\
    MCP & $100.22\pm \phantom{0}5.84$  &  $77.60\pm19.50$  & $0.19\pm0.03$ & $1420.60\pm97.30$  &$139.40\pm42.15$ & $0.07\pm0.01$\\
    \fos\     & $\phantom{00}0.12\pm \phantom{0}0.08$       & $\phantom{0}1.00\pm\phantom{0}2.82$  & $0.19\pm0.04$ & $\phantom{000}5.72\pm \phantom{0}2.69$      & $\phantom{00}0.00\pm\phantom{0}0.00$ & $0.10\pm0.02$ \\
    \bottomrule
  \end{tabular}
\end{table*}

\begin{table*}
\caption{Average run times (in seconds), Hamming distances, and Estimation errors for \grlassocv\ and \grfos\ for data that vary in  group length (GL) and number of activated groups (AG).
For the larger data set, \grlassocv\  timed out on our machine, which means that \grlassocv\  took more than one hour for the first run.
The results illustrate that \grfos\ is both faster and almost more accurate than  \grlassocv\ across all settings } 
\label{tab:time_hammingGroup}
\centering
  \begin{tabular}{c c c c c c c}
 
    \toprule 
      & \multicolumn{5}{c}{$~\nobs=500,\,p=1000$} \\ 
    & \multicolumn{2}{c}{$GL=5,\,AG=1$} & \multicolumn{4}{c}{$GL=5,\,AG=40$}                  \\
	\cline{2-7}
    Method     & Timing     & Hamming distance & Estimation error & Timing     & Hamming distance & Estimation error\\
    \grlassocv     & $259.25\pm 43.03$ & $133.50\pm 47.15$  &$0.53\pm0.08$  & $333.87\pm 29.95$ & $671.50\pm 109.31$ & $0.20\pm0.02$
     \\

    \grfos    & $\phantom{00}5.14\pm \phantom{0}0.81$       & $\phantom{00}0.00\pm\phantom{0}0.00$  &$0.14\pm0.03$   & $\phantom{00}2.88\pm \phantom{0}0.93$ & $195.50\pm \phantom{00}5.50$ &$0.21\pm 0.02$  \\
        \toprule
        & \multicolumn{5}{c}{$~\nobs=500,\,p=1000$}\\
    & \multicolumn{2}{c}{$GL=10,\,AG=2$} & \multicolumn{4}{c}{$GL=10,\,AG=10$}  \\
    \cline{2-7}
    Method     & Timing     & Hamming distance & Estimation error & Timing     & Hamming distance & Estimation error\\
    
   \grlassocv     & $282.35\pm41.34$ & $216.00\pm 112.17$  &$0.31\pm0.06$    & $322.19\pm33.04$ & $520.00\pm87.55$ &$0.23\pm0.02$    \\

    \grfos     & $\phantom{00}3.06\pm\phantom{0}0.70$       & $\phantom{00}0.00\pm\phantom{00}0.00$ &$0.14\pm0.01$   & $\phantom{00}2.08\pm \phantom{0}0.45$      & $\phantom{0}32.00\pm41.04$  &$0.18\pm0.01$  \\
     \toprule
    & \multicolumn{5}{c}{$~\nobs=5000,\,p=10\,000$}\\
    & \multicolumn{2}{c}{$GL=10,\,AG=2$} & \multicolumn{4}{c}{$GL=10,\,AG=10$}                  \\
	\cline{2-7}
    Method     & Timing     & Hamming distance & Estimation error & Timing     & Hamming distance & Estimation error\\
   \grlassocv     & NA & NA & NA    & NA & NA  & NA  \\

    \grfos     & $\phantom{0}258.19\pm 153.28$       & $2.00\pm\phantom{}6.32$ &$0.05\pm0.02$   & $124.97\pm 36.74$ & $\phantom{}2.00\pm \phantom{}6.32$ &$0.07\pm 0.01$   \\
    \bottomrule
    
  \end{tabular}

\end{table*}
We summarize the results for feature selection and group-feature selection  in Table~\ref{tab:time_hamming} and Table~\ref{tab:time_hammingGroup}, respectively.  
For each setting, that is, for each a combination of $n$, $p$, group length, and number of activated groups, we recorded the timings of each method.
The computational efficiency is measured in average timing (in seconds) over 10 independent runs;
the statistical accuracy is measured in average Hamming distance, which is the sum of the number of false positives and the number of false negatives, and in average estimation error, which is $\normsup{\estimatorfinal-\target}$.
If a method timed out on our machines, we put an ``NA.'' 
We observe that \fos\ outperforms \lcvs, \lcvg, SCAD, and MCP and that \grfos\ outperforms \grlassocv\ both in computational efficiency and in statistical accuracy.
For large data sets, \lcvs\  and \grlassocv\  timed out on our machine, which means that the run time for the first run  is one hour or more. 
We also compared \fos\  with AV approach~\citep[Algorithm~1]{Chichignoud_Lederer_Wainwright14} in Table~\ref{tab:Liner_CHI} (of Appendix~\ref{app:B2}).

The two computational benefits of our proposed method are illustrated in Fig.~\ref{fig:iterations}. First, we observe that even with warm starts, \lcvs\ requires a large number of iterations to converge. In contrast, \fos\ allows for early stopping, in particular, for large tuning parameters (recall that the required precision for \fos\ is proportional to the tuning parameter; instead, the required precision for other methods is unknown). Moreover, Cross-Validation, BIC, AIC, and similar calibration schemes are based on the entire \lasso\ path, while only a part of the path is required for \fos. We should remark that the same result holds for \grfos.

\begin{figure}[!ht]
  \centering
\begin{tikzpicture}[scale=0.75]
\begin{semilogyaxis}[
	xlabel=$-\log_{10}(\tuningparameter)$,
	ylabel=Number of iterations,
	grid=major,
	legend pos=south east]

\addplot[densely dashed, color=blue, line width = 2pt] coordinates {
    (-3.160030e+00,0)
(-3.129727e+00,100)
(-3.099424e+00,100)
(-3.069121e+00,100)
(-3.038818e+00,100)
(-3.008515e+00,100)
(-2.978212e+00,100)
(-2.947909e+00,100)
(-2.917606e+00,100)
(-2.887303e+00,100)
(-2.857000e+00,100)
(-2.826696e+00,100)
(-2.796393e+00,100)
(-2.766090e+00,100)
(-2.735787e+00,100)
(-2.705484e+00,100)
(-2.675181e+00,100)
(-2.644878e+00,100)
(-2.614575e+00,100)
(-2.584272e+00,100)
(-2.553969e+00,100)
(-2.523666e+00,100)
(-2.493363e+00,100)
(-2.463060e+00,100)
(-2.432757e+00,100)
(-2.402454e+00,100)
(-2.372151e+00,100)
(-2.341848e+00,100)
(-2.311545e+00,100)
(-2.281242e+00,100)
(-2.250939e+00,100)
(-2.220636e+00,100)
(-2.190333e+00,100)
(-2.160030e+00,100)
(-2.129727e+00,100)
(-2.099424e+00,100)
(-2.069121e+00,100)
(-2.038818e+00,100)
(-2.008515e+00,100)
(-1.978212e+00,100)
(-1.947909e+00,100)
(-1.917606e+00,100)
(-1.887303e+00,100)
(-1.857000e+00,100)
(-1.826696e+00,100)
(-1.796393e+00,100)
(-1.766090e+00,100)
(-1.735787e+00,100)
(-1.705484e+00,100)
(-1.675181e+00,100)
(-1.644878e+00,100)
(-1.614575e+00,200)
(-1.584272e+00,200)
(-1.553969e+00,200)
(-1.523666e+00,200)
(-1.493363e+00,300)
(-1.463060e+00,300)
(-1.432757e+00,300)
(-1.402454e+00,300)
(-1.372151e+00,200)
(-1.341848e+00,200)
(-1.311545e+00,300)
(-1.281242e+00,300)
(-1.250939e+00,400)
(-1.220636e+00,300)
(-1.190333e+00,400)
(-1.160030e+00,400)
(-1.129727e+00,400)
(-1.099424e+00,500)
(-1.069121e+00,600)
(-1.038818e+00,500)
(-1.008515e+00,400)
(-9.782116e-01,600)
(-9.479086e-01,600)
(-9.176056e-01,700)
(-8.873025e-01,700)
(-8.569995e-01,500)
(-8.266965e-01,700)
(-7.963935e-01,500)
(-7.660904e-01,700)
(-7.357874e-01,700)
(-7.054844e-01,700)
(-6.751813e-01,700)
(-6.448783e-01,1001)
(-6.145753e-01,1001)
(-5.842722e-01,900)
(-5.539692e-01,1000)
(-5.236662e-01,1000)
(-4.933632e-01,1000)
(-4.630601e-01,1000)
(-4.327571e-01,900)
(-4.024541e-01,1001)
(-3.721510e-01,900)
(-3.418480e-01,1001)
(-3.115450e-01,1000)
(-2.812419e-01,1001)
(-2.509389e-01,1000)
(-2.206359e-01,1000)
(-1.903328e-01,1001)
(-1.600298e-01,1001)
};

\addplot[color=red, line width = 2pt] coordinates {
(-3.010526e+00,0)
(-2.980223e+00,1)
(-2.949920e+00,1)
(-2.919617e+00,1)
(-2.889314e+00,1)
(-2.859011e+00,1)
(-2.828708e+00,1)
(-2.798405e+00,1)
(-2.768102e+00,1)
(-2.737799e+00,1)
(-2.707496e+00,1)
(-2.677193e+00,1)
(-2.646890e+00,1)
(-2.616587e+00,1)
(-2.586284e+00,1)
(-2.555981e+00,1)
(-2.525678e+00,1)
(-2.495375e+00,1)
(-2.465072e+00,1)
(-2.434769e+00,1)
(-2.404466e+00,1)
(-2.374163e+00,1)
(-2.343860e+00,1)
(-2.313556e+00,1)
(-2.283253e+00,1)
(-2.252950e+00,1)
(-2.222647e+00,1)
(-2.192344e+00,1)
(-2.162041e+00,1)
(-2.131738e+00,1)
(-2.101435e+00,1)
(-2.071132e+00,1)
(-2.040829e+00,1)
(-2.010526e+00,1)
(-1.980223e+00,1)
(-1.949920e+00,1)
(-1.919617e+00,1)
(-1.889314e+00,1)
(-1.859011e+00,1)
(-1.828708e+00,1)
(-1.798405e+00,1)
(-1.768102e+00,1)
(-1.737799e+00,1)
(-1.707496e+00,1)
(-1.677193e+00,1)
(-1.646890e+00,1)
(-1.616587e+00,1)
(-1.586284e+00,1)
(-1.555981e+00,1)
(-1.525678e+00,5)
(-1.495375e+00,2)
(-1.465072e+00,25)
(-1.434769e+00,11)
(-1.404466e+00,4)
(-1.374163e+00,3)
(-1.343860e+00,34)

};
\end{semilogyaxis}
\end{tikzpicture}
\vspace{-.8\baselineskip}
\caption{The red, solid line depicts the number of proximal gradient steps in \emph{FOS}  as a function of the tuning parameter~$\tuningparameter$.  
The blue, dashed line depicts the corresponding number of proximal gradient steps in \lcvs. 
Shown are the numbers for one data set in the $\nobs=500$ and $p=1000$ setting.  
We observe that \lcvs\ requires a large number of iterations to converge, while \fos\ allows for early stopping, in particular, for large tuning parameters. 
Also, \lcvs\ is based on the entire \lasso\ path, while only a part of the path is required for \fos.\label{fig:iterations}}
\end{figure}

\paragraph{Financial data}\label{finance}
Now, we consider a large data set to demonstrate the scalability of \fos. 
The data~\citep{Kogan09} comprises~$\nobs=16\,087$ samples and~$p=150\,348$ predictors. 
 The goal is to use financial reports of companies to predict the volatility of stock returns. 
The feature representation of the financial reports is based on the calculation of TF-IDF (term
frequency and inverse document frequency) of unigrams. 
There is no ground truth available for verifying statistical accuracies,
but the data is ideal for verifying the algorithm's scalability.
The computational time of \lcvg\ is {\bf $\mathbf{124.56}$s} and of  \fos\ {\bf $\mathbf{1.94}$s}. In contrast, \lcvs, SCAD, and MCP timed out on our machine,
which again means more than one hour of computation. 
This shows that our methodology is indeed highly scalable to large data.

\paragraph{Lung cancer data set}\label{Bio}
\fos\ can also be applied to network learning problems by estimating the local neighborhood of each node via high-dimensional regressions. In this specific application, the goal is to understand the interaction network of~$p=1000$ genes in lung cancer patients from~$\nobs=500$ expression profiles~\citep{Guyon08}. We do neighborhood selection with the ``or-rule''~\citep{Meinshausen06} based on \fos\ and lassoCV and compare the estimated graphs with the available gold standard~\citep{Statnikov15}. The results are summarized in Fig.~\ref{fig:reged}. Note that here, the Hamming distance is the sum of the falsely included edges and the falsely omitted edges.
We find that our pipeline outmatches the standard one both in  speed and accuracy.\vspace{3mm}
\paragraph{Breast cancer data set}\label{breastcancer}
Here, we consider a large data set that  contains gene expressions from 60~patients with estrogen-positive breast cancer \citep{Xiao}.
The patients were treated for five years and then classified into two categories (labeled by $1$ or $-1$) according to whether the cancer recurred or not. 
The original data is pre-processed as follows: first, the genes with more than~$50$ percent missingness are removed and all other missing values are filled  by mean imputation. This  reduces  number  of  genes  from  $22\,575$  to  $12\,071$. Then, the genes are  grouped  using cytogenetic position data, namely the C1~set from the GSEA method~\citep{GSEA}.
The genes that are not recorded in the C1~set are removed, which yields a total of $4989$~genes in
270~groups, with an average group size of 18.5 genes.
Finally, the data is split via $10$-fold Cross-Validation  and \grlassocv\ and \grfos\ are applied. 
The classification is then performed by taking the signum function on the predicted values.
We report the classification errors in Fig.~\ref{fig:Breastcancer}. 
We find that \grfos\ outperforms \grlassocv\ in computational efficiency and  accuracy.
\begin{figure*}[!ht]

\begin{minipage}[c]{.45\linewidth}
\begin{bchart}[min=0, max=3000, step=1000, width=0.9\linewidth]
\bcbar[text={\lcvg},color=orange,plain]{2551.8}
\bcbar[text={\fos},color=blue!25,plain]{238.02}
\bcxlabel{{\scriptsize Run time (seconds)}}
\end{bchart}

\end{minipage}\hfill
\begin{minipage}[c]{.5\linewidth}
\begin{bchart}[min=0, max=4, step=1, width=0.8\linewidth]

\bcbar[text={\lcvg},color=orange,plain]{3.3}
\bcbar[text={\fos},color=blue!25,plain]{0.45}

\bcxlabel{{\scriptsize Hamming distance (\% of total number of possible edges)}}
\end{bchart}

\end{minipage}
\caption{Run times (in seconds) and Hamming distances (in \% of the total number of possible edges) for \lcvg\ and for \fos\ on the lung cancer data set. The implementations \lcvs, SCAD, and MCP timed out on our machine.
The results illustrate that \fos\ is both faster and more accurate than  \lcvg.  \label{fig:reged}}
 \vspace{0.5\baselineskip}
\end{figure*}
\begin{figure*}[!ht]

\begin{minipage}[c]{.45\linewidth}
\begin{bchart}[min=0, max=300, step=50, width=0.9\linewidth]
\bcbar[text={\grlassocv},color=orange,plain]{260.90}
\bcbar[text={\grfos},color=blue!25,plain]{19.16}
\bcxlabel{{\scriptsize Run time (seconds)}}
\end{bchart}
\end{minipage}\hfill
\begin{minipage}[c]{.5\linewidth}
\begin{bchart}[min=0, max=30, step=5, width=0.8\linewidth]
 
\bcbar[text={\grlassocv},color=orange,plain]{26}
\bcbar[text={\grfos},color=blue!25,plain]{22}

\bcxlabel{{\scriptsize Classification error }}
\end{bchart}

\end{minipage}
\caption{Run times (in seconds) and classification error for  \grlassocv\ and \grfos\ on the breast cancer data set
 illustrate that \grfos\ is both faster and more accurate than \grlassocv.
  \label{fig:Breastcancer}}
 \vspace{0.5\baselineskip}
\end{figure*}

\subsection{Logistic Regression}\label{logreg}
Here, we specify our approach to
 $\ell_1$-regularized  logistic regression. 
 In Section~\ref{Methodlog}, we provide  sufficient conditions for Assumption~\eqref{assumptionstat} (see Lemma~\ref{lemma:logregstat}). 
 In Section~\ref{Alglog}, we specify the implementation for feature selection (see Algorithm~\ref{logregalgo}), and in Section~\ref{numericslogreg}, we demonstrate the empirical efficiency of our approach both on simulated and real data.
\subsubsection{Method}\label{Methodlog}

The logistic regression model is
\begin{linenomath}
	\begin{equation*}\label{logmodel}
          \text{Pr}(y_i=1\,|\, \observation_i) = \frac{\exp(\observation_i\tp\target)}{1+\exp(\observation_i\tp\target)}~~~~~~~~i\in\otn\,,
	\end{equation*}
\end{linenomath}
where~$\outcome=(y_1,\ldots,y_n)\tp\in\{0,1\}^n$ is a binary vector outcome, $\observation_1,\ldots,\observation_n$ are the rows of the design matrix $\design\in\Rnp$, and $\target\in\Rp=:\parameterset$  the regression vector with support~$\targetset=\supp(\target)\subset\otp$. 

To estimate $\target$, we consider the objective function
\begin{linenomath}
	\begin{equation}\label{logregobj}
	\objective(\parameter,\tuningparameter):=L(\parameter)+\tuningparameter\normone{\parameter}\,,
	\end{equation}
\end{linenomath}
where~$\parameter\in\Rp$, $\tuningparameter\in[0,\infty)=:\tuningparameterset$, and $L(\parameter):=(\sum_{i=1}^n\log(1+\exp(\observation_i\tp\parameter))-y_i\observation_i\tp\parameter)/n$. 
We recall that the $\ell_1$-regularization allows us to control the $\ell_\infty$-distance of our estimators.   
This corresponds in our framework to the measure of accuracy $\distance(\parameter,\parameter') :=\normsup{\parameter-\parameter'}$.
To avoid digression, we focus on $\ell_1$-norm regularization; 
it is straightforward to extend the results in this section to more general classes of regularizers as in Section~\ref{LR}.
We then propose to use the estimators~$\estimatorset=\bigl\{\estimatora:\tuningparameter\in\tuningparameterset\bigr\}$ with
\begin{linenomath}
\begin{equation}\label{RLogRegobj}
  \estimatora\in\argmin_{\parameter\in\Rp}\objective(\parameter,\tuningparameter)\,.
\end{equation}
\end{linenomath}

\nce{irrepresentability}{\gamma}

For vectors $\boldsymbol{u},\boldsymbol{v}$ of the same length, we define the function $\functionw(\boldsymbol{u},\boldsymbol{v}) := \exp(\boldsymbol{u}\tp \boldsymbol{v})/(1+\exp(\boldsymbol{u}\tp \boldsymbol{v}))^2$ and
we denote by $\matrixw$ the diagonal matrix of size $n\times n$ with $(i,i)$th entry $\functionw(\observation_i,\target)$.
The residuals are defined as $u_i := y_i-\text{Pr}(y_i=1\,|\, \observation_i)$ for $i\in\otn$. 
A sufficient condition for satisfying the statistical bound in Assumption~\eqref{assumptionstat}  for estimators in Equation~\eqref{RLogRegobj}  is  provided by~\citet[Theorem~1]{Li2019} as follows:
\begin{lemma}[\functionstat\ in  $\ell_1$-regularized logistic regression]\label{lemma:logregstat}
	Assume
	\begin{linenomath}
	\begin{equation}\label{EVCond}
	\cminlogreg:=\Omega_{\min}(\design_\targetset\tp\matrixw\design_\targetset/n)>0\,
	\end{equation} 
\end{linenomath}
	and
	\begin{linenomath}
	\begin{equation}\label{IReCond}
	\irLog := 1 - \normsupmat{(\design_\targetset\tp\matrixw\design_\targetset)^{-1}\design_\targetset\tp\matrixw\design_{\targetseti}}>0\,.
	\end{equation} 
	\end{linenomath}
	Then,~\estimatorset\ satisfies
	\begin{linenomath}
	\begin{equation*}
	\normsup{\target-\estimatora} \leq \functionstata~~~\forall\, \tuningparameter\geq \tuningparametero
	\end{equation*}
	\end{linenomath}
	for~$\functionstata:=\constantlogreg\tuningparameter$ with $\constantlogreg:=1.5\normsupmat{(\design_\targetset\tp\matrixw\design_\targetset)^{-1}}/(\normtwomat{(\design_\targetset\tp\matrixw\design_{\targetset})^{-1}}\cminlogreg)$, and~$\tuningparametero:=4(2-\irLog )\normsup{\design\tp\noise}/(n\irLog )$.
\end{lemma}
\noindent This result implies that for a suitable tuning   parameter~\tuningparameter,
the estimator \estimatora\ is uniformly close to the regression vector \target. 
Condition~\eqref{EVCond}  is a modified version of the minimal eigenvalue condition commonly used in the theory for linear regression~\citep[Equation~(11.29)]{Hastie2015},~\citep[Assumption~1]{Li2019} and ensures that the relevant covariates are only mildly correlated.
 Condition~\eqref{IReCond} is a modified version of irrepresentability condition commonly used in the theories about the lasso~\citep[Equation~(11.27)]{Hastie2015},~\citep[Assumption~2]{Li2019},~\citep{Zhao06} and prevents the relevant covariates from being strongly correlated with the irrelevant covariates. 
Note that these assumptions are slightly different to (and potentially more stringent than) the assumptions in the linear-regression case;
it would be interesting to try relaxing these assumptions---especially in view of our simulations showing that our proposed method performs well even for data with larger correlations (see Appendix~\ref{app:B2})---but this is beyond the scope of the current paper.

A sufficient condition for satisfying the computational bound in Assumption~\eqref{assumptioncomp}  for the estimators in Equation~\eqref{RLogRegobj} follows again from Lemma~\ref{combound}.

Plugging Lemma~\ref{lemma:logregstat}  and Lemma~\ref{combound} into Theorem~\ref{mainresult} yields the following result.
\begin{corollary}[Estimation in  $\ell_1$-regularized logistic regression]\label{EsLOGR}
 Suppose that the conditions of Lemma~\ref{lemma:logregstat} are met and 
   $\functioncomp=\functionstat$. Then, the outputs of Algorithm~\ref{mainalgorithm} satisfy
   \begin{linenomath}
  \begin{equation*}
  \tuningparameterestimatorfinal\leq \tuningparametero, 
\end{equation*}
\end{linenomath}
\begin{linenomath}
 \begin{equation*}
    \normsup{\estimatorfinal-\target}\leq 6\constantlogreg\tuningparametero,
  \end{equation*}
  \end{linenomath}
and
\begin{linenomath}
 \begin{equation*}
    \E[ \normsup{\estimatorfinal-\target}]\leq \E[6\constantlogreg\tuningparametero].
  \end{equation*}
\end{linenomath}
\end{corollary}
This equips the proposed algorithm  with guarantees for its estimation error in logistic regression.

An estimate of the support is
\begin{linenomath}
\begin{equation}\label{Log-reg-thresholding}
\estimatorS\,:=\,\bigl\{j\in\otp\,:\,|\estimatorfinal_j|> 6\constantlogreg\tuningparameterestimatorfinal \bigr\}\,,
\end{equation}
\end{linenomath}
where~\tuningparameterestimatorfinal\ and~\estimatorfinal\ are the outputs of Algorithm~\ref{mainalgorithm}. 
Now, assuming the conditions of Lemma~\ref{lemma:logregstat}  and again a  standard beta-min condition
\begin{linenomath}
\begin{equation*}
  \min_{j\in\targetset}|\target_j|>  12\constantlogreg\tuningparametero,
\end{equation*}
\end{linenomath}
we obtain a guarantee for~$\estimatorS$.

  \begin{corollary}[Feature selection in $\ell_1$-regularized logistic regression]\label{log-reg-FS}
Under the stated conditions, it holds that
\begin{linenomath}
\begin{equation*}
  \estimatorS\,=\,\targetset\,.
\end{equation*}
\end{linenomath}
  \end{corollary}
Hence, 
the feature selection scheme in~\eqref{Log-reg-thresholding} is equipped with a sharp guarantee that takes both the tuning parameter calibration and the computational tolerances of implementations into account.

\subsubsection{Algorithm}\label{Alglog}
 Algorithm~\ref{logregalgo}  is a specification of  our general scheme for  feature selection in $\ell_1$-regularized logistic regression.
We call the algorithm \logfos\ for convenient reference. 
The three challenges described earlier  are now addressed as follows: (1)~The stopping point on the tuning parameter path is determined via AV-tests. (2)~To ensure that the computational bound is reached, we follow the same approach as in linear regression case (see Section~\ref{lassoalgo}). Setting~$\functionstata:=g(\tuningparameter):=\constantlogreg\tuningparameter$, which by Lemma~\ref{combound}, gives the stopping criterion as $\dgap{\estimatorc}{\cd}\le \dif:=\re\nobs\tuningparameter^2(\constantlogreg-(1/\re\nobs))^2$.
(3)~We finally use Equation~\eqref{Log-reg-thresholding}  for thresholding, which guarantees correct  feature selection (see Corollary~\ref{log-reg-FS}).   
As initialization, we choose the all-zeros vector in~$\Rp,$ reflecting our assumption that many features are inactive. 
Since we are limited to finitely many computations in practice, we consider finite sequences~$r_1=r_{\max}>r_2>\dots>\tuningparameter_M=\tuningparameter_{\min}>0.$ The concrete choice follows the ones used in~\citet[Section~3]{Li2019}. We consider $M=500$ tuning parameters that are equally spaced on $[\tuningparameter_{\min},\tuningparameter_{\max}]$, where $\tuningparameter_{\min}:=0.0001\tuningparameter_{\max}$ and $\tuningparameter_{\max}:=10\log(p)/n$ ensure a large spread of outcomes.
Finally,
$\constantlogreg=6$ and $\re=1$ are considered global constants, see~\citet[Section~3]{Li2019} for more details regarding $\constantlogreg$. 
We show later that changing these constants has minor impact on our results (see Tables~\ref{Const_impact_Log_r} and~\ref{C_impact_Log_r}).

\begin{algorithm}[!ht]\vspace{1mm}
	\SetKwInOut{Input}{Inputs}\SetKwInOut{Output}{Outputs}
	\Input{$\outcome\in\{0,1\}^n$; $\design\in \Rnp$; $\tuningparameter_1=\tuningparameter_{\max} > \tuningparameter_2 > \ldots > \tuningparameter_M>0$}
	\Output{$\estimatorc^{\tuningparameterestimatorfinal}\in\Rp$; $\estimatorS\subset\otp$}
	\BlankLine
	\textbf{Initialization :} \texttt{statsCont:=true}; \texttt{statsIt:=}1; $\estimatorc^{\tuningparameter_1}:=\zerof{p}$; $\tuningparameterestimatorfinal:=\tuningparameter_M$\; 
	\While{\texttt{statsCont==true} \texttt{AND} $\texttt{\textup{statsIt}}\,<M$}{  
		\texttt{statsIt:=statsIt+1}\;
		\texttt{compCont:=true}\;
		\texttt{betaOld:=}$\estimatorc^{\tuningparameter_{\texttt{\textup{statsIt}}-1}}$\;
		\While{\texttt{compCont==true}}{
			 Compute a feasible dual point $\cd^{\tuningparameter_\texttt{\textup{statsIt}}}$ \; 
        Compute the duality gap $\dgap{\estimatorc^{\tuningparameter_\texttt{\textup{statsIt}}}}{\cd^{\tuningparameter_\texttt{\textup{statsIt}}}}$\;
        \eIf{$\dgap{\estimatorc^{\tuningparameter_\texttt{\textup{statsIt}}}}{\cd^{\tuningparameter_\texttt{\textup{statsIt}}}}\leq \,\dif$}{
				$\estimatorc^{\tuningparameter_\texttt{\textup{statsIt}}}$\texttt{:=betaOld}\;
				\texttt{compCont:=false}\;
			}{
				$\estimatorc^{\tuningparameter_\texttt{\textup{statsIt}}}:=$one iteration of Solver applied to \texttt{betaOld}\;
				\texttt{betaOld:=}$\estimatorc^{\tuningparameter_\texttt{\textup{statsIt}}}$\;
			}
		}
		\texttt{statsCont:=}$\displaystyle \prod_{i=1}^{\texttt{\textup{statsIt}}} \mathds{1}\bigl\{ \normsup{\estimatorc^{\tuningparameter_\texttt{\textup{statsIt}}} - \estimatorc^{\tuningparameter_i}}/(\tuningparameter_\texttt{\textup{statsIt}} + \tuningparameter_i) - 2\constantlogreg \leq 0\bigr\}$\;
	}
	\If{\texttt{statsCont==false}}{
		$\tuningparameterestimatorfinal:=\tuningparameter_{\texttt{\textup{statsIt}}-1}$\;
	}
	$\estimatorS\,:=\,\bigl\{j\in\otp:|\estimatorfinal_j|> 6\constantlogreg\tuningparameterestimatorfinal\bigr\}$
	\caption{feature selection in $\ell_1$-regularized logistic regression}\label{logregalgo}
\end{algorithm}

\subsubsection{Empirical performance}\label{numericslogreg}
We now demonstrate the computational efficiency and the empirical accuracy of the proposed method on simulated and real data.   
We compare \logfos\ to the \loglasso\ with Cross-Validation (log-lassoCV).  log-lassoCV is implemented analogously to \logfos\ using the \texttt{SPAMS} package and a 10-fold Cross-Validation with warm starts. 
We call this implementation $\loglcvs$.

\paragraph{Synthetic data}\label{SyntheticdataLogreg}
We first  use synthetic data to demonstrate the empirical performance of \logfos.
Data are generated from the logistic regression model with~$\nobs=200$ and~$p=200$,~$\nobs=200$ and~$p=500$, and also~$\nobs=1000$ and~$p=5000$.
Each row of the design matrix~$\design\in\Rnp$ is sampled independently from a~$p$-dimensional normal distribution with mean~$0$ and covariance matrix~$(1-\rho)\matrixI+\rho\1,$ where~$\matrixI$ is the identity matrix, $\1$  the matrix of ones, and~$\rho\in\{0.25,0.5\}$  the population correlation among the features. 
The design matrix is then normalized such that its columns have Euclidean norm equal to~$\sqrt n.$ 
The entries of~$\target$ are first set to~$0$ except for~$s=8$
uniformly at random chosen entries that are set to $1$ or $-1$ with equal probability. 
The computational efficiency is measured in average timing (in seconds) over 10 independent runs; 
the statistical accuracy is measured in average Hamming distance and in average estimation error, which is $\normsup{\estimatorfinal-\target}$.
Table~\ref{tab:time_hamming-log-reglargedata} reports the results for  data generated  with different size and correlation. We find that \logfos\ outperforms \loglcvs\ in computational efficiency and accuracy even for data with high correlation. 
The results for the other correlation and sparsity levels are deferred to the Appendix~\ref{app:B2}.
We also compared \logfos\  with~\citet{Li2019} in Table~\ref{tab:log-reg-LI} (of Appendix~\ref{app:B2}).
 
 \begin{table*}
\centering
\caption{Average run times (in seconds), Hamming distances, and Estimation errors for \loglcvs\  and  \logfos, with $M=500$.  
For large data set, \loglcvs\ timed out on our machine, which means that \loglcvs\ took more than one hour for the first run. 
The results illustrate that \logfos\ is both faster and more accurate than  \loglcvs\   across all settings} 
\label{tab:time_hamming-log-reglargedata}
  \begin{tabular}{c c  c c c c c}
 
    \toprule 
      & \multicolumn{5}{c}{$~\nobs=200,\,p=200,s=8$} \\ 
    & \multicolumn{2}{c}{$\rho=0.25$} & \multicolumn{4}{c}{$\rho=0.5$}                  \\
	\cline{2-7}
    Method     & Timing     & Hamming distance &Estimation error & Timing     & Hamming distance & Estimation error \\
     \loglcvs\     & $13.96\pm 0.59$       & $7.80\pm3.04$ &$2.89\pm0.57$    & $15.77\pm 0.33$       & $10.80\pm2.29$ &$2.96\pm0.57$   \\

    \logfos    & $\phantom{0}0.36\pm0.06$       & $3.10\pm1.79$  &$2.04\pm0.59$  & $\phantom{0}0.58\pm 0.03$       & $\phantom{0}5.60\pm2.11$  &$2.35\pm0.69$ \\
        \toprule
        & \multicolumn{5}{c}{$~\nobs=200,\,p=500,s=8$}\\
    & \multicolumn{2}{c}{$\rho=0.25$} & \multicolumn{4}{c}{$\rho=0.5$}  \\
    \cline{2-7}
    Method     & Timing     & Hamming distance & Estimation error & Timing     & Hamming distance  & Estimation error \\
    
     \loglcvs\     & $26.12\pm 0.58$      & $11.10\pm4.38$  &$3.17\pm0.20$   & $26.48\pm 0.66$      & $13.90\pm3.87$  &$3.32\pm0.21$\\

    \logfos     & $\phantom{0}1.22\pm0.07$      & $\phantom{0}4.40\pm1.95$  &$2.36\pm 0.16$  & $\phantom{0}1.09\pm 0.04$      & $\phantom{0}8.60\pm3.83$ &$2.56\pm0.36$ \\
     \toprule
    & \multicolumn{5}{c}{$~\nobs=1000,\,p=5000,s=8$}\\
    & \multicolumn{2}{c}{$\rho=0.25$} & \multicolumn{4}{c}{$\rho=0.5$}                  \\
	\cline{2-7}
    Method     & Timing     & Hamming distance & Estimation error  & Timing     & Hamming distance & Estimation error\\
     \loglcvs\     & NA & NA  & NA  & NA & NA  & NA \\

    \logfos     & $33.64\pm4.69$       & $\phantom{0}4.70\pm1.33$  & $2.48\pm0.58$ & $25.95\pm 4.23$      & $\phantom{0}3.80\pm1.68$  & $2.54\pm 0.57$ \\
    \bottomrule
    
  \end{tabular}

\end{table*}
\paragraph{Melanoma patients}
We compare the performance of \logfos\ against  \loglcvs\ classifying melanoma patients~\citep{mian2005serum}, including
 $n = 205$ mass spectrometry scans of serum samples collected from $101$ patients with early-stage melanoma  and $104$ patients with advanced stage melanoma. 
 Each scan measures the intensities for $18\,856$ mass over charge (m/Z) values. 
Finding  m/Z values  associated with the stage of the disease is of our interest for classifying data.
We fit a logistic regression model  with the  same data pre-processing  step as in~\citet{lederer2015don}: 
We  start with a peak filtering algorithm to extract  $p = 500$ most relevant m/Z values. 
We also fill  $\outcome\in \{-1,1\}^{205}$ with
$y_{i} = -1$ for $i = 1,\dots,101$ (early-stage) and $y_{i} = 1$ for $i = 102,\dots,205$ (advanced stage).
 Finally, we use $10$-fold Cross-Validation  splitting data and \loglcvs\ and \logfos\ are applied.  Fig.~\ref{fig:Melanoma} reports the mean 10-fold CV classification errors  by  \logfos\ and  \loglcvs, as well as their run times. We find that \logfos\ outperforms \loglcvs\
 in computational efficiency and accuracy.
\begin{figure*}[!ht]

\begin{minipage}[c]{.45\linewidth}
\begin{bchart}[min=0, max=100, step=20, width=0.9\linewidth]
\bcbar[text={\loglcvs},color=orange,plain]{90.92}
\bcbar[text={\logfos},color=blue!25,plain]{1.78}
\bcxlabel{{\scriptsize Run time (seconds)}}
\end{bchart}
\end{minipage}\hfill
\begin{minipage}[c]{.5\linewidth}
\begin{bchart}[steps={0.1,0.2,0.3},max=0.3]
 
\bcbar[text={\loglcvs},color=orange,plain]{0.2536}
\bcbar[text={\logfos},color=blue!25,plain]{0.1505}

\bcxlabel{{\scriptsize Classification error }}
\end{bchart}

\end{minipage}
\caption{Run times (in seconds) and classification error for  \loglcvs\ and \logfos\ on the melanoma patients illustrate that \logfos\ is both faster and more accurate than  \loglcvs.
  \label{fig:Melanoma}}
 \vspace{0.5\baselineskip}
\end{figure*}

\section{Discussion}\label{discussion}

In view of the theoretical and empirical evidence provided above, 
the presented approach provides competitive methods for feature  selection and group-feature  selection with large and high-dimensional data. 
In particular, there are no comparable theoretical guarantees that provide a connection between statistical and computational accuracy,
and in our simulations and real data applications, the algorithm rivals or outmatches its competitors in terms of speed and accuracy.

Our theories do not discuss the optimal choice of the set of tuning parameters over which to optimize;
this might be a direction for future research.

\appendices

\section{Proofs}\label{app:appendixa}

We provide proofs of our main results and derive one property of vectors that are close to a solution $\estimatora$ of~\eqref{lassoobj} and~\eqref{RLogRegobj}: 
in Lemma~\ref{lemma:cone}, we show that their error belongs to a cone.
\\
 \begin{proof}[Theorem~\ref{mainresult}]
We prove the three claims in order.\\
\emph{Claim 1:} We first prove that~$\tuningparameterestimatorfinal\leq\tuningparametero$. 
We show this by contradiction and assume that~$\tuningparameterestimatorfinal>\tuningparametero.$ 
By definition of~\tuningparameterestimatorfinal, this means that there are tuning parameters~$\tuningparameter',\tuningparameter''\geq \tuningparametero$ such  that 
\begin{linenomath}
\begin{equation*}
    \distance(\estimatorc^{\tuningparameter'},\estimatorc^{\tuningparameter''})> \functionstat(\tuningparameter')+\functioncomp(\tuningparameter')+\functionstat(\tuningparameter'')+\functioncomp(\tuningparameter'')\,.
\end{equation*}
\end{linenomath}
On the other hand, by the assumed symmetry and triangle inequality of~\distance, it holds that
\begin{linenomath}
\begin{equation*}
    \distance(\estimatorc^{\tuningparameter'},\estimatorc^{\tuningparameter''})\leq \distance(\estimatorc^{\tuningparameter'},\target) + \distance(\estimatorc^{\tuningparameter''},\target)\,.
\end{equation*}
\end{linenomath}
Using the symmetry and triangle inequality of~\distance\ again and combining with Assumptions~\eqref{assumptionstat}  and~\eqref{assumptioncomp} (statistical and computational bounds) yields for the first term
\begin{linenomath}
\begin{equation*}
    \distance(\estimatorc^{\tuningparameter'},\target)\leq \distance(\estimator^{\tuningparameter'},\target)+\distance(\estimatorc^{\tuningparameter'},\estimator^{\tuningparameter'}) \leq \functionstat(\tuningparameter')+\functioncomp(\tuningparameter')\,,
  \end{equation*}
  \end{linenomath}
and similarly for the second term
\begin{linenomath}
 \begin{equation*}
    \distance(\estimatorc^{\tuningparameter''},\target)\leq \functionstat(\tuningparameter'')+\functioncomp(\tuningparameter'')\,.
  \end{equation*}
  \end{linenomath}
It follows that
\begin{linenomath}
\begin{equation*}
    \distance(\estimatorc^{\tuningparameter'},\estimatorc^{\tuningparameter''})\leq \functionstat(\tuningparameter')+\functioncomp(\tuningparameter')+\functionstat(\tuningparameter'')+\functioncomp(\tuningparameter'')\,,
\end{equation*}
\end{linenomath}
which contradicts the initial display, and thus concludes the proof of the first inequality.

\emph{Claim 2:} We now prove that
\begin{linenomath}
 \begin{equation*}
    \distance(\estimatorfinal,\target)\leq 3\functionstatao+3\functioncompao\,.
  \end{equation*}
  \end{linenomath}
For this, we first use the symmetry and triangle inequality of~\distance\ to find
\begin{linenomath}
\begin{equation*}
\distance(\estimatorfinal,\target)\leq \distance(\estimatorc^{\tuningparametero},\target) + \distance(\estimatorc^{\tuningparametero},\estimatorfinal)\,.    
\end{equation*}
\end{linenomath}
The first term can be bounded similarly as in Claim~1. 
We find
\begin{linenomath}
\begin{equation*}
\distance(\estimatorc^{\tuningparametero},\target)\leq \functionstatao+\functioncompao\,.    
\end{equation*}
\end{linenomath}
The second term can be bounded by virtue of the definition of the test and~$\tuningparameterestimatorfinal\leq \tuningparametero$ according to the Claim~1. We find
\begin{linenomath}
\begin{equation*}
\distance(\estimatorc^{\tuningparametero},\estimatorfinal)\leq  \functionstatao+\functioncompao+\functionstat(\tuningparameterestimatorfinal)+\functioncomp(\tuningparameterestimatorfinal)\,.    
\end{equation*}
\end{linenomath}
We can now combine the terms and use that~$\tuningparameterestimatorfinal\leq \tuningparametero$ and that $\functionstat,\functioncomp$ are increasing to find
\begin{linenomath}
\begin{equation*}
\distance(\estimatorfinal,\target)\leq 3\functionstatao+3\functioncompao,
\end{equation*}
\end{linenomath}
as desired.\\
\emph{Claim 3:} We finally prove that
\begin{linenomath}
\begin{equation*}
\E[\distance(\estimatorfinal,\target)]\leq 3\E\functionstatao+3\E\functioncompao\,.
\end{equation*}
\end{linenomath}
This follows directly from Claim~2 by taking expectations.
\end{proof}
\begin{lemma}[Relaxed guarantee for~$\tuningparameterestimatorfinal$ \ and~$\estimatorfinal$]
\label{pracresult}
Under Assumption~\ref{feasibility}, any estimator~$\estimatorca$,~$\tuningparameter\geq \tuningparameterestimatorfinal$~with~$\tuningparameterestimatorfinal$ from Algorithm~\ref{mainalgorithm}, satisfies
\begin{linenomath}
 \begin{equation*}
    \distance(\estimatorca,\target)\leq 3\functionstat(\max\{\tuningparametero,\tuningparameter\})+3\functioncomp(\max\{\tuningparametero,\tuningparameter\})\,.
  \end{equation*}
  \end{linenomath}
\end{lemma}
\begin{proof}[Lemma~\ref{pracresult}]

By the symmetry and triangle inequality of~\distance,
\begin{linenomath}
\begin{equation*}
      \distance(\estimatorca,\target)\leq \distance(\estimatorc^{\tuningparametero},\target) + \distance(\estimatorc^{\tuningparametero},\estimatorca)\,. 
\end{equation*}
\end{linenomath}
The first term can be bounded as in the earlier proof:
\begin{linenomath}
\begin{equation*}
\distance(\estimatorc^{\tuningparametero},\target) \leq \functionstatao+\functioncompao\,.    
\end{equation*}
\end{linenomath}
For the second term, we use that~$\tuningparameterestimatorfinal\leq\tuningparameter,\tuningparametero$ by assumption/earlier proof and the tests in Algorithm~\ref{mainalgorithm} to deduce that
\begin{linenomath}
\begin{equation*}
\distance(\estimatorc^{\tuningparametero},\estimatorca)\leq  \functionstatao+\functioncompao+\functionstat(\tuningparameter)+\functioncomp(\tuningparameter)\,.    
\end{equation*}
\end{linenomath}
Collecting the pieces and using and that~$\functionstat,\functioncomp$ are increasing yields
\begin{linenomath}
\begin{equation*}
      \distance(\estimatorca,\target)\leq 3\functionstat(\max\{\tuningparametero, \tuningparameter\})+3\functioncomp(\max\{\tuningparametero, \tuningparameter\})\,,
\end{equation*}
\end{linenomath}
as desired.
\end{proof}
\begin{proof}[Lemma~\ref{groupres}]
	We will first prove that~$\estimatorSi\subset \targetseti$. 
	For this, consider~$j\in\estimatorSi$ and note that by the triangle inequality
	\begin{linenomath}
	\begin{multline*}
	\norm{\target_{\groupj}}_q\,=\,\norm{\target_{\groupj}-\estimatorfinal_{\groupj}+\estimatorfinal_{\groupj}}_q \\
	 \leq \norm{\target_{\groupj}-\estimatorfinal_{\groupj}}_q+\norm{\estimatorfinal_{\groupj}}_q\,.
	\end{multline*}
	\end{linenomath}
	The first term can be bounded by using Theorem~\ref{mainresult}, Result~\eqref{r2}:
	\begin{linenomath}
	\begin{equation*}
	\norm{\target_{\groupj}-\estimatorfinal_{\groupj}}_q\leq \norm{\target-\estimatorfinal}_{q,\infty}\leq 3\functionstatao+3\functioncompao\,.
	\end{equation*}
	\end{linenomath}
	The second term can be bounded by using~$j\in\estimatorSi$ and the definition of \estimatorS: 
	\begin{linenomath}
	\begin{equation*}
	\norm{\estimatorfinal_{\groupj}}_q\leq 3\functionstat(\tuningparameterestimatorfinal)+3\functioncomp(\tuningparameterestimatorfinal)\,.
	\end{equation*}
	\end{linenomath}
	Collecting the pieces and using that~$\tuningparameterestimatorfinal\leq \tuningparametero$ (Theorem~\ref{mainresult}, Result~\eqref{r1}) gives
	\begin{linenomath}
	\begin{multline*}
	\norm{\target_{\groupj}}_q \leq 3\functionstatao+3\functioncompao+3\functionstatao+3\functioncompao\\
	= 6\functionstatao+6\functioncompao\,,
	\end{multline*}
	\end{linenomath}
	which means by virtue of the beta-min condition that~$j\in\targetseti$.
	This concludes the proof of the first part.

	Next, we prove that~$\targetseti\subset\estimatorSi$. Using~$j\in\targetseti$, that is~$\target_{\groupj}=\zerof{\groupsizej}$, we obtain
	\begin{linenomath}
	\begin{equation*}
	\norm{\estimatorfinal_{\groupj}}_q = \norm{\estimatorfinal_{\groupj}-\target_{\groupj}}_q \leq 3\functionstatao+3\functioncompao\,,
	\end{equation*}
	\end{linenomath}
	where the inequality is obtained similarly to the first part of the proof.
\end{proof}
\newcommand{\SubDif}{\boldsymbol{\kappa'}}
\begin{lemma}[Subdifferential of regularizer]\label{SubDif_lamma}
For all  $\SubDif\in \partial\normt{\parameter'}$, where $\partial\normt{\parameter'} $ is  the subdifferential of the regularizer at  $\parameter'\in\Rp$, we have $\norm{\SubDif}\le 1$ (recall that $\norm{\cdot}$ is the dual norm of~$\normt{\cdot}$).
\end{lemma}

\begin{proof}[Lemma~\ref{SubDif_lamma}]
We prove that $\norm{\SubDif}\le 1$ for all $\SubDif\in \partial\normt{\parameter'}$.

Since $\SubDif\in \partial\normt{\parameter'}$, by definition we have for all $\parameter \in \Rp$
\begin{linenomath}
\begin{equation*}
\normt{\parameter}\ge \normt{\parameter'}+\inprod{\SubDif}{\parameter-\parameter'}
\end{equation*}
\end{linenomath}
and by some rearranging we have
\begin{linenomath}
\begin{equation*}
\inprod{\SubDif}{\parameter'}-\normt{\parameter'}\ge  \inprod{\SubDif}{\parameter}-\normt{\parameter}\,.
\end{equation*}
\end{linenomath}
Because above inequality holds for all $\parameter\in \Rp$, we also obtain
\begin{linenomath}
\begin{equation*}
\inprod{\SubDif}{\parameter'}-\normt{\parameter'}\ge \sup_{\parameter\in\Rp} \bigl\{\inprod{\SubDif}{\parameter}-\normt{\parameter}\bigr\}\,.
\end{equation*}
\end{linenomath}
The right hand side of the inequality above represents the  Fenchel conjugate of $\normt{\cdot}$ at $\SubDif$. We use~\citet[Proposition~1.4]{bach2011optimization} to obtain 
\begin{linenomath}
\begin{multline*}
\inprod{\SubDif}{\parameter'}-\normt{\parameter'}\ge \sup_{\parameter\in\Rp} \bigl\{\inprod{\SubDif}{\parameter}-\normt{\parameter}\bigr\}\\
=
\begin{cases}
0 & \norm{\SubDif}\le 1\,;\\
+\infty & \text{otherwise\,.}
\end{cases}
\end{multline*}
\end{linenomath}
The left-hand side of the display is finite for each given $\parameter',\SubDif$;
hence,
it must hold that $\norm{\SubDif}\le 1$.

\end{proof}
\begin{proof}[Lemma~\ref{linreg}]
We prove the result in two steps.\\
\emph{Step 1:} We first prove that $\Ldifdelta:=\target-\estimatora$ satisfies 
\begin{linenomath}
\begin{equation*}
    \normt{\Ldifdelta_{\targetseti}}\leq 3\normt{\Ldifdelta_{\targetset}}\,.
\end{equation*}
\end{linenomath}
By definition of \estimatora, we find the basic inequality
\begin{linenomath}
\begin{equation*}
  \frac{1}{2}\normtwo{\outcome-\design\target}^2+\tuningparameter\normt{\target}\geq \frac{1}{2}\normtwo{\outcome-\design\estimatora}^2+\tuningparameter\normt{\estimatora}.
\end{equation*}
\end{linenomath}
We can now rewrite $\normtwo{\outcome-\design\estimatora}^2/2$ as follows:
\begin{linenomath}
\begin{align*}
    \frac{1}{2}\normtwo{\outcome&-\design\estimatora}^2 \\
    &= \frac{1}{2}\normtwo{\outcome-\design\target+\design\target-\design\estimatora}^2\\
    &= \frac{1}{2}\normtwo{\outcome-\design\target}^2 + \inprod{\outcome-\design\target}{\design\target-\design\estimatora}\\
    &~~~~+ \frac{1}{2}\normtwo{\design\target-\design\estimatora}^2\\
    &= \frac{1}{2}\normtwo{\outcome-\design\target}^2 + \inprod{\design\tp \left(\outcome-\design\target\right)}{\target-\estimatora}\\
    &~~~~+ \frac{1}{2}\normtwo{\design\target-\design\estimatora}^2\,.
\end{align*}
\end{linenomath}
Combining the two displays yields
\begin{linenomath}
\begin{multline*}
\tuningparameter\normt{\target}\geq \inprod{\design\tp \left(\outcome-\design\target\right)}{\target-\estimatora}+ \frac{1}{2}\normtwo{\design\target-\design\estimatora}^2\\ +\tuningparameter\normt{\estimatora}\,.
\end{multline*}
\end{linenomath}
We now use that $\normtwo{\design\target-\design\estimatora}^2/2\geq 0$ and invoke the model $\outcome=\design\target+\noise$ to find
\begin{linenomath}
\begin{equation*}
\tuningparameter\normt{\target}\geq \inprod{\design\tp\noise}{\target-\estimatora}+\tuningparameter\normt{\estimatora}\,.
\end{equation*}
\end{linenomath}
This can be rearranged to
\begin{linenomath}
\begin{equation*}
\tuningparameter\normt{\estimatora}\leq \inprod{\design\tp\noise}{\estimatora-\target}+\tuningparameter\normt{\target}\,.
\end{equation*}
\end{linenomath}
Invoking  H\"older's inequality and the assumption that $\tuningparameter\geq 2\norm{\design\tp\noise}$ provides us with
\begin{linenomath}
\begin{align*}
  \inprod{\design\tp\noise}{\estimatora-\target}&\leq \norm{\design\tp\noise}\normt{\estimatora-\target}\\
  &\leq \frac{\tuningparameter}{2}\normt{\estimatora-\target}\,.
\end{align*}
\end{linenomath}
Together with the above inequality, we thus find
\begin{linenomath}
\begin{equation*}
\tuningparameter\normt{\estimatora}\leq \frac{\tuningparameter}{2}\normt{\estimatora-\target}+\tuningparameter\normt{\target}\,.
\end{equation*}
\end{linenomath}
We then divide both sides by $r\in(0,\infty)$ and find
\begin{linenomath}
\begin{equation*}
\normt{\estimatora}\leq \frac{1}{2}\normt{\estimatora-\target}+\normt{\target}\,.
\end{equation*} 
\end{linenomath}
According to the assumed decomposability of $\normt{\cdot}$ with respect to $\targetset$, we can now decompose each of these vectors into their parts on \targetset\ and \targetseti:
\begin{linenomath}
\begin{align*}
 \normt{\estimatoratargetset}+\normt{\estimatoratargetseti}&\leq\frac{1}{2}\normt{\estimatoratargetset-\targettargetset}+\frac{1}{2}\normt{\estimatoratargetseti-\targettargetseti}\\
 &~~~~+\normt{\targettargetset}+\normt{\targettargetseti}\\
 &= \frac{1}{2}\normt{\estimatoratargetset-\targettargetset}+\frac{1}{2}\normt{\estimatoratargetseti}+\normt{\targettargetset}\,.
\end{align*}
\end{linenomath}
We can rearrange the terms and find
\begin{linenomath}
\begin{equation*}
    \frac{1}{2}\normt{\estimatoratargetseti}\leq \frac{1}{2}\normt{\estimatoratargetset-\targettargetset}+\normt{\targettargetset}-\normt{\estimatoratargetset}\,.
\end{equation*}
\end{linenomath}
Using the reverse triangle inequality, this becomes
\begin{linenomath}
\begin{equation*}
    \frac{1}{2}\normt{\estimatoratargetseti}\leq \frac{3}{2}\normt{\estimatoratargetset-\targettargetset}\,. 
\end{equation*}
\end{linenomath}
Finally, setting $\Ldifdelta:=\target-\estimatora$ and multiplying both sides by $2$ gives the desired inequality
\begin{linenomath}
\begin{equation*}
    \normt{\Ldifdelta_{\targetseti}}\leq 3\normt{\Ldifdelta_{\targetset}}\,. 
\end{equation*}
\end{linenomath}
\emph{Step 2:} We now prove that \estimatorset\ satisfies
\begin{linenomath}
  \begin{equation*}
     \norm{\target-\estimatora}\leq \frac{3\tuningparameter}{2\nobs\compatibility}~~~\text{for all }\tuningparameter\geq 2\norm{\design\tp\noise}\,.
 \end{equation*}
 \end{linenomath}
For this, we first observe that the KKT conditions for the objective function at \estimatora\ are
\begin{linenomath}
\begin{equation*}
  -\design\tp(\outcome-\design\estimatora)+\tuningparameter\vsub = \zerof{p}\,,
\end{equation*}
\end{linenomath}
where $\vsub\in\partial\normt{\estimatora}$ (where $\partial\normt{\estimatora} $ is  the subdifferential of regularizer at  $\estimatora$). 
Using the model, we have
\begin{linenomath}
\begin{equation*}
  -\design\tp(\design\target+\noise-\design\estimatora)+\tuningparameter\vsub=\zerof{p}\,,
\end{equation*}
\end{linenomath}
and rearranging yields
\begin{linenomath}
\begin{equation*}
     \design\tp  X(\target-\estimatora) = \tuningparameter\vsub - \design\tp\noise\,.
\end{equation*}
\end{linenomath}
Applying the norm on both sides and using the linearity and triangle inequality for norms gives
\begin{linenomath}
\begin{align*}
    \norm{\design\tp  X(\target-\estimatora)} &= \norm{\tuningparameter\vsub - {\design\tp\noise}}\\
    &\leq \tuningparameter\norm{\vsub} + \norm{\design\tp\noise}\,.
\end{align*}
\end{linenomath}
Using now that $\norm{\vsub}\leq 1$ ($\vsub\in\partial\normt{\estimatora}$ and we use our Lemma~\ref{SubDif_lamma}) and $\tuningparameter\geq 2\norm{\design\tp\noise}$ (by assumption) gives
\begin{linenomath}
\begin{equation*}
    \norm{\design\tp  X(\target-\estimatora)}\leq \tuningparameter + \frac{\tuningparameter}{2}=\frac{3\tuningparameter}{2}\,.
\end{equation*}
\end{linenomath}
Now, we recall that  $\Ldifdelta:=\target-\estimatora$ satisfies $\normt{\Ldifdelta_{\targetseti}} \leq 3\normt{\Ldifdelta_{\targetset}}$ according to Step~1. Thus, the assumption given in the lemma entails $\norm{\target-\estimatora}\leq \norm{\design\tp \design(\target-\estimatora)}/(\nobs\compatibility)$, so that
\begin{linenomath}
\begin{equation*}
    \norm{\target-\estimatora}\leq \frac{3\tuningparameter}{2\nobs\compatibility},
\end{equation*}
\end{linenomath}
as desired.
\end{proof}

\begin{lemma}[Cone constraint in regularized  regression]\label{lemma:cone}
	Let $\diflog\in[0,\infty)$ be a constant and $\estimatorc\in\Rp$ any vector that satisfies $\objective(\estimatorc,2\tuningparameter)\leq \objective(\estimatora,2\tuningparameter)+\diflog$. Then $\difdelta:=\estimatorc-\estimatora$ belongs to the cone
	\begin{linenomath}
	\begin{equation*}
	\cones:= \left\{\vdiff\in\Rp\,:\,\normt{\vdiff_{\targetseti}}\leq \coneparam\normt{\vdiff_{\targetset}}+4\normt{\estimatora_{\targetseti}} + \frac{\diflog}{\tuningparameter} \right\}\,.
	\end{equation*}
	\end{linenomath}
\end{lemma}
\begin{proof}[Lemma~\ref{lemma:cone}]
Since $\objective(\estimatorc,2\tuningparameter)\leq \objective(\estimatora,2\tuningparameter)+\diflog$\,, we find the basic inequality
\begin{linenomath}
\begin{equation*}
L(\estimatora+\difdelta)+2\tuningparameter\normt{\estimatora+\difdelta}\le L(\estimatora)+2\tuningparameter\normt{\estimatora}+\diflog\,.
\end{equation*}
\end{linenomath}
 This inequality is equivalent to
 \begin{linenomath}
\begin{equation*}
 L(\estimatora+\difdelta)- L(\estimatora)+2\tuningparameter\normt{\estimatora+\difdelta}-2\tuningparameter\normt{\estimatora}\le\diflog\,.
\end{equation*}
\end{linenomath}
Next, we find a lower bound for the left-hand side of the above display in two steps: \\
\emph{Step 1:} First, we show that 
\begin{linenomath}
\begin{equation*}
	L(\estimatora+\difdelta)-L(\estimatora) \geq -\tuningparameter(\normt{\difdelta_\targetset}+\normt{\difdelta_{\targetseti}})\,.
\end{equation*}
\end{linenomath}
By convexity of the loss function $L$, we have
\begin{linenomath}
\begin{equation*}
	L(\estimatora+\difdelta)-L(\estimatora)\geq \inprod{\nabla L(\estimatora)}{\difdelta}\geq -|\inprod{\nabla L(\estimatora)}{\difdelta}|\,.
\end{equation*}
\end{linenomath}
The KKT condition for $\estimatora$ provides us with $\nabla L(\estimatora)=-\tuningparameter\vsub$, for some vector $\vsub\in\partial\normt{\estimatora}$.  
Invoking H\"older's inequality, applying the decomposability of $\normt{\cdot}$ with respect to $\targetset$, and using that $\norm{\vsub}\le 1$ (by Lemma~\ref{SubDif_lamma})  yields
\begin{linenomath}
\begin{multline*}
	L(\estimatora+\difdelta)-L(\estimatora)\geq -\tuningparameter\norm{\vsub}(\normt{\difdelta_\targetset}+\normt{\difdelta_{\targetseti}})\\
	\geq-\tuningparameter(\normt{\difdelta_\targetset}+\normt{\difdelta_{\targetseti}})\,.
\end{multline*}
\end{linenomath}
\emph{Step 2:} Now we prove that 
\begin{linenomath}
\begin{equation*}
	\normt{\estimatora+\difdelta}-\normt{\estimatora}\geq \normt{\difdelta_{\targetseti}}-2\normt{\estimatora_{\targetseti}}-\normt{\difdelta_\targetset}\,.
\end{equation*}
\end{linenomath}
Using that $\normt{\estimatora+\difdelta}=\normt{\estimatora_{\targetset}+\estimatora_{\targetseti}+\difdelta_{\targetset}+\difdelta_{\targetseti}}$ and applying the triangle inequality gives
\begin{linenomath}
\begin{equation*}
	\normt{\estimatora+\difdelta}\geq \normt{\estimatora_{\targetset}+\difdelta_{\targetseti}}-\normt{\estimatora_{\targetseti}+\difdelta_{\targetset}}\,.
\end{equation*}
\end{linenomath}
Subtracting $\normt{\estimatora}$  from both sides of the above inequality and using decomposability condition, we obtain
\begin{linenomath}
\begin{align*}
	\normt{\estimatora&+\difdelta}-\normt{\estimatora}\\
	&\geq
	\normt{\estimatora_{\targetset}+\difdelta_{\targetseti}}-\normt{\estimatora_{\targetseti}+\difdelta_{\targetset}}-\normt{\estimatora}\\ 
	&=\normt{(\estimatora_{\targetset}+\difdelta_{\targetseti})_{\targetset}}+
	\normt{(\estimatora_{\targetset}+\difdelta_{\targetseti})_{\targetseti}}\\
	&~~
	-\normt{(\estimatora_{\targetseti}+\difdelta_{\targetset})_{\targetset}}
	-\normt{(\estimatora_{\targetseti}+\difdelta_{\targetset})_{\targetseti}}
	-\normt{\estimatora}\\	
	&=\normt{\estimatora_{\targetset}}+\normt{\difdelta_{\targetseti}}-\normt{\estimatora_{\targetseti}}-\normt{\difdelta_{\targetset}}\\
	&~~-\normt{\estimatora_{\targetset}}-\normt{\estimatora_{\targetseti}}\\
	&=\normt{\difdelta_{\targetseti}}-2\normt{\estimatora_{\targetseti}}-\normt{\difdelta_{\targetset}}\,.
\end{align*}
\end{linenomath}
Combining the results of Step~1 and Step~2 gives
\begin{linenomath}
\begin{align*}
    L(\estimatora+\difdelta)&- L(\estimatora)+2\tuningparameter\normt{\estimatora+\difdelta}-2\tuningparameter\normt{\estimatora} \\
    &\geq -\tuningparameter(\normt{\difdelta_\targetset}+\normt{\difdelta_{\targetseti}}) + 2\tuningparameter\normt{\difdelta_{\targetseti}}\\
    &~~~-4\tuningparameter\normt{\estimatora_{\targetseti}}-2\tuningparameter\normt{\difdelta_\targetset}\\
    &= \tuningparameter\normt{\difdelta_{\targetseti}}-3\tuningparameter\normt{\difdelta_\targetset}-4\tuningparameter\normt{\estimatora_{\targetseti}} \,.
\end{align*}
\end{linenomath}
Finally, plugging this into the basic inequality and rearranging, we find
\begin{linenomath}
\begin{equation*}
    \normt{\difdelta_{\targetseti}}\leq 3\normt{\difdelta_\targetset}+ 4\normt{\estimatora_{\targetseti}}+\frac{\diflog}{\tuningparameter},
\end{equation*}
\end{linenomath}
as desired.
\end{proof}
\begin{proof}[Lemma~\ref{combound}]
	Since $\objective(\estimatorc,2\tuningparameter)\leq  \objective(\estimatora,2\tuningparameter)+\dif,\,$ we find the basic inequality
	\begin{linenomath}
	\begin{equation*}
	L(\estimatorc)+2\tuningparameter\normt{\estimatorc} \le L(\estimatora)+2\tuningparameter\normt{\estimatora}+\dif\,.
	\end{equation*}
	\end{linenomath}
	We now recall that the KKT conditions read
	\begin{linenomath}
	\begin{equation*}
	\nabla L(\estimatora) + \tuningparameter\vsub = \zerof{p}\,,
	\end{equation*}
	\end{linenomath}
	for some vector $\vsub\in\partial\normt{\estimatora}$.
	Adding a null term to the right-hand side of the previous display gives
	\begin{linenomath}
	\begin{multline*}
	L(\estimatorc)+2\tuningparameter\normt{\estimatorc} \leq L(\estimatora)+2\tuningparameter\normt{\estimatora}+\dif \\
	+ \inprod{\nabla L(\estimatora) + \tuningparameter\vsub}{\estimatorc-\estimatora}\,.
	\end{multline*}
	\end{linenomath}
	Rearranging yields
	\begin{linenomath}
	\begin{multline*}
	L(\estimatorc)-L(\estimatora) - \inprod{\nabla L(\estimatora)}{\estimatorc-\estimatora} \\
	\leq \inprod{\tuningparameter\vsub}{\estimatorc-\estimatora} -2\tuningparameter\normt{\estimatorc} + 2\tuningparameter\normt{\estimatora}+\dif \,.
	\end{multline*}
	\end{linenomath}
	Using that  $\norm{\vsub}\leq 1$ (Lemma~\ref{SubDif_lamma}) and $\vsub\tp\estimatora = \normt{\estimatora}$, we get
	\begin{linenomath}
	\begin{multline*}
	L(\estimatorc)-L(\estimatora) - \inprod{\nabla L(\estimatora)}{\estimatorc-\estimatora} \\
	\leq  \tuningparameter(\vsub\tp\estimatorc-\normt{\estimatorc})-\tuningparameter\normt{\estimatorc} + \tuningparameter\normt{\estimatora}+\dif\,,
	\end{multline*}
	\end{linenomath}
	and by H\"older's inequality, $\vsub^\top\estimatorc\leq \norm{\vsub}\normt{\estimatorc} \leq \normt{\estimatorc}$\,.
	Therefore,
	\begin{linenomath}
	\begin{equation*}
	L(\estimatorc)-L(\estimatora) - \inprod{\nabla L(\estimatora)}{\estimatorc-\estimatora} \leq \tuningparameter\normt{\estimatora} -\tuningparameter\normt{\estimatorc} + \dif\,,
	\end{equation*}
	\end{linenomath}
	so that with the reverse triangle inequality
	\begin{linenomath}
	\begin{align*}
	L(\estimatorc)-L(\estimatora) - \inprod{\nabla L(\estimatora)}{\estimatorc-\estimatora} &\leq \tuningparameter\normt{\estimatorc-\estimatora} + \dif\,.
	\end{align*}
	\end{linenomath}
	We finally get, setting $\difdelta:=\estimatorc-\estimatora,$
	\begin{linenomath}
	\begin{equation*}
	L(\estimatora+\difdelta)-L(\estimatora) - \inprod{\nabla L(\estimatora)}{\difdelta} \leq \tuningparameter\normt{\difdelta} + \dif\,.
	\end{equation*}
	\end{linenomath}
	Moreover, Lemma~\ref{lemma:cone} guarantees that $\difdelta\in\cones$ and thus allows us to apply the condition in the upper display of the lemma. 
	So we also find 
	\begin{linenomath}
	\begin{equation*}
	L(\estimatora+\difdelta)-L(\estimatora)- \inprod{\nabla L(\estimatora)}{\difdelta}\geq \re\nobs{\normt{\difdelta}}^2 \,.
	\end{equation*}
	\end{linenomath}
	Combining these two inequalities gives us
	\begin{linenomath}
	\begin{equation*}
	\re\nobs{\normt{\difdelta}}^2 \leq \tuningparameter\normt{\difdelta}+\dif\,.
	\end{equation*}
	\end{linenomath}
	Dividing both sides of the inequality by $\re\nobs>0$, adding $\tuningparameter^2/(2\re\nobs)^2$, and rearranging gives
	\begin{linenomath}
	\begin{equation*}
	{\normt{\difdelta}}^2 - \frac{\tuningparameter}{\re\nobs}\normt{\difdelta}+\Bigl(\frac{\tuningparameter}{2\re\nobs}\Bigr)^2\leq \frac{\dif}{\re\nobs}+\Bigl(\frac{\tuningparameter}{2\re\nobs}\Bigr)^2\,.
	\end{equation*}
	\end{linenomath}
	Rewriting the previous display yields
	\begin{linenomath}
	\begin{equation*}
	\Bigl(\normt{\difdelta}-\frac{\tuningparameter}{2\re\nobs}\Bigr)^2\leq \frac{\dif}{\re\nobs}+\Bigl(\frac{\tuningparameter}{2\re\nobs}\Bigr)^2	
	\end{equation*}
	\end{linenomath}
	and
	\begin{linenomath}
	\begin{equation*}
	\normt{\difdelta}\leq \sqrt{\frac{\dif}{\re\nobs}+\Bigl(\frac{\tuningparameter}{2\re\nobs}\Bigr)^2}+\frac{\tuningparameter}{2\re\nobs}\,.
	\end{equation*}
	\end{linenomath}
	Finally, using the assumption that $\norm{\cdot}\le \normt{\cdot}$, we obtain
	\begin{linenomath}
	\begin{equation*}
	\norm{\difdelta}\leq \sqrt{\frac{\dif}{\re\nobs}+\Bigl(\frac{\tuningparameter}{2\re\nobs}\Bigr)^2}+\frac{\tuningparameter}{2\re\nobs}\le \sqrt{\frac{\dif}{\re\nobs}}+\frac{\tuningparameter}{\re\nobs}\,,
	\end{equation*}
	\end{linenomath}
	as desired.
\end{proof}

\section{Additional Materials}\label{app:appendixb}
We finally provide additional theoretical and empirical results.
\subsection{On the Condition of Lemma~\ref{combound}}\label{sec:condition}

We have stated in the main part of the paper that the condition in Lemma~\ref{combound} is a slightly stronger version of restricted strong convexity.
Here, we corroborate that statement.
We specify the setup of Section~\ref{LR} a bit further:
we (i)~assume Gaussian noise ($\noise\sim\mathcal{N}[\boldsymbol{0},\sigma^2 \matrixI]$),
(ii)~impose slightly more restrictive assumptions on the design (the irrepresentability condition \citep[Assumption~7.3.1]{LedererBook}),
(iii)~assume $\ell_1$-regularization ($\norm{\cdot}=\normsup{\cdot}$),
and (iv)~redefine the optimal tuning parameter slightly ($\tuningparametero:=4\normsup{\design\tp\noise}$). 
These additional specifications facilitate the derivations and help us to focus on the main ideas.

Assumption~(iii) allows us to use the primal-dual witness approach as detailed in \citet[Section~7.3]{LedererBook} to show that $\normt{\estimator^{\tuningparameter}_{\targetseti}}=0$ for all $\tuningparameter\geq \tuningparametero$. 
Thus, the condition in the lemma simplifies to
\begin{linenomath}
	\begin{equation*}
	L(\parameter+\vdiff)-L(\parameter)-\inprod{\nabla L(\parameter)}{\vdiff}=\normtwo{\design\vdiff}^2\geq \re\nobs\norm{\vdiff}_1^2 
	\end{equation*}
	\end{linenomath}
	for all $\parameter\in\Rp$ and $\vdiff\in\Rp$ that satisfy $\normone{\vdiff_{\targetseti}}\leq \coneparam\normone{\vdiff_{\targetset}} + \dif/\tuningparameter$.
We now distinguish two cases:

\emph{Case 1: $\normone{\vdiff_{\targetset}} \geq \dif/\tuningparameter$}

Since
\begin{multline*}
  \norm{\vdiff}_1\leq \norm{\vdiff_{\targetset}}_1+\norm{\vdiff_{\targetseti}}_1\leq \norm{\vdiff_{\targetset}}_1+\coneparam\normone{\vdiff_{\targetset}} + \dif/\tuningparameter \\ \leq 5\normone{\vdiff_{\targetset}}\leq 
  5\sqrt{\abs{\targetset}}\normtwo{\vdiff_{\targetset}}\leq 5\sqrt{\abs{\targetset}}\normtwo{\vdiff}\,,
\end{multline*}
the condition stated in the lemma  becomes (note that using the argument above, we are considering the worst case of our curvature assumption)
\begin{linenomath}
	\begin{equation*}
	\normtwo{\design\vdiff}^2\geq 25\re\nobs{\abs{\targetset}}\normtwo{\vdiff}^2\,,
	\end{equation*}
	\end{linenomath}
which equals the usual restricted strong convexity up to a rescaling of the constant $\re$ by $25{\abs{\targetset}}$ \citep[Pages~291ff]{Hastie2015}.
(Whether that rescaling can be avoided or is intrinsic to the problem of connecting statistical and computational bounds would be an interesting topic for further research.)

\emph{Case 2: $\normone{\vdiff_{\targetset}} < \dif/\tuningparameter$}

We find using 
1.~the assumed decomposability of the $\ell_1$-norm, 
2.~the ``cone condition'' for~$\vdiff$,
and 3.~the assumption $\normone{\vdiff_{\targetset}} < \dif/\tuningparameter$ that
\begin{align*}
  \normone{\vdiff}&=\normone{\vdiff_{\targetset}}+\normone{\vdiff_{\targetseti}}\\
  &\leq\normone{\vdiff_{\targetset}}+ \coneparam\normone{\vdiff_{\targetset}} + \dif/\tuningparameter\\
&\leq 5\dif/\tuningparameter\,.
\end{align*}
Now, since $\vdiff:=\estimatorc-\estimatora$ satisfies $\normt{\vdiff_{\targetseti}}\leq \coneparam\normt{\vdiff_{\targetset}} + \dif/\tuningparameter$ according to Lemma~\ref{lemma:cone},
we find $\normone{\estimatorc-\estimatora}\leq 5\dif/\tuningparameter$ for any vector $\estimatorc\in\Rp$ that satisfies $\objective(\estimatorc,2\tuningparameter)\leq \objective(\estimatora,2\tuningparameter)+\diflog$.
Hence,
with our choice $\dif=\tuningparameter^2/(zn)$,
we find a suitable computational bound (cf.~Lemma~\ref{linreg})
\begin{linenomath}
  \begin{equation*}
    \distance(\estimatorca,\estimatora)=\normsup{\estimatorc-\estimatora}\leq 5\tuningparameter/(zn)~~~~~~\forall\, \tuningparameter\in\tuningparameterset
  \end{equation*}
  \end{linenomath}
without any restrictions altogether.

The two cases combined illustrate that the condition in Lemma~\ref{combound} is stronger but still comparable to restricted strong convexity.
It would be interesting to study questions of optimality in this context,
but this is beyond the scope of this paper.


\subsection{Additional Optimization Theory}\label{app:B1}
Our approach requires the construction of a feasible dual point for the approximated estimator~\estimatorc.
In the following, 
we describe how to find an explicit expression for such a point.
Observe that a dual formulation  of the primal objective~\eqref{lassoobj} is~\citep{Borwein2010,Ndiaye16}
\begin{linenomath}
\begin{multline*}
\td:= \argmax_{\dualpoint\in\Rn} D(\dualpoint,\tuningparameter):=-\frac{\tuningparameter^2}{2}\normtwo{\dualpoint-\frac{\outcome}{\tuningparameter}}^2+\frac{1}{2}\normtwo{\outcome}^2\\
\text{subject to}~\norm{\design^\top\dualpoint} \leq 1\,.
\end{multline*}
\end{linenomath}
A primal solution of~\eqref{lassoobj} and the dual solution~\td\ above are linked by 
\begin{linenomath}
\begin{equation*}
    \design\estimatora =  \outcome-\tuningparameter\td\,.
\end{equation*}
\end{linenomath}
This suggests choosing as feasible dual point a rescaled version of the residuals. 
Given a current primal estimate $\estimatorc$, this yields $\cd= s (\outcome-\design\estimatorc)$, where $s$ is given by~\citep{Ghaoui2010}
\begin{linenomath}
\begin{multline*}
s := \min\biggl\{\max\biggl\{\frac{-1}{\norm{\design^\top(\outcome-\design\estimatorc)}},\frac{\outcome^\top(\outcome-\design\estimatorc)}{\tuningparameter\normtwo{\outcome-\design\estimatorc}^2}\biggr\},\\
\frac{1}{\norm{\design^\top(\outcome-\design\estimatorc)}}\biggr\}\,.
\end{multline*}
\end{linenomath}
This choice for the coefficient $s$ ensures that \cd\ is the closest (in $\ell_2$-norm) point to $\outcome/\tuningparameter$ in the feasible set $\{\dualpoint\in\Rn\,:\,\norm{\design^\top\dualpoint} \leq 1\}$.

\subsection{Additional Simulations}\label{app:B2}
In this section, 
we give additional simulation results for the settings of Sections~\ref{synthetic} and~\ref{SyntheticdataLogreg}. 

Table~\ref{tab:time_hamming-largeM} contains results for the setting of Section~\ref{synthetic} with
varying sizes~$M$ of the tuning parameter set. 
The results show that all methods are robust with respect to the specific choice of the tuning parameter set.

Table~\ref{tab:time_hamming-log-adddata} contains results for the  setting of Section~\ref{SyntheticdataLogreg} with
varying correlation~$\rho$. The results show that our method also works for correlated data.

Table~\ref{tab:Liner_CHI} contains results for the  setting of Section~\ref{synthetic} to compare  \fos\ with~\citet[Algorithm~1]{Chichignoud_Lederer_Wainwright14} (we call it \CHICHI), which corresponds to running \fos\ with $b$ close to zero (we set $b=1e-7$).
The results suggest that \fos\ is much faster and more accurate than \CHICHI.

Table~\ref{tab:log-reg-LI} contains results for the  setting of Section~\ref{SyntheticdataLogreg} to compare  \logfos\ with~\citet{Li2019} (we call it \LI), which corresponds to running  \logfos\  with $b$ close to zero (we set again $b=1e-7$). 
The results suggest that  \logfos\  is much faster  than~\LI\ in all settings and similarly accurate.

Tables~\ref{Const_impact_LR} and~\ref{C_impact_LR} contain results for the  setting of Section~\ref{synthetic} with
varying constants $\re$ and $\compatibility$, respectively. 
The results show that changing $\re$ and $\compatibility$ has minor impact on our results.

Tables~\ref{Const_impact_Log_r} and~\ref{C_impact_Log_r} contain results for the  setting of Section~\ref{SyntheticdataLogreg} with
varying constants $\re$ and $\constantlogreg$, respectively. 
The results show that changing $\re$ and $\constantlogreg$ has minor impact on our results.
 \begin{table*}
\centering
\caption{the average run times (in seconds) and average Hamming distances in these settings
illustrate the limited influence of the specific choice of the tuning parameter set}
\label{tab:time_hamming-largeM}
  \begin{tabular}{c c c c c}
 
    \toprule 
      & \multicolumn{4}{c}{$~\nobs=500,\,p=500,s=10,\rho=0.3$} \\ 
    & \multicolumn{2}{c}{$M=500$} & \multicolumn{2}{c}{$M=1000$}                  \\
	\cline{2-5}
    Method     & Timing     & Hamming distance & Timing     & Hamming distance\\
     \lcvs\     & $313.34\pm 21.93$       & $50.80\pm16.07$    & $578.00\pm 43.15$       & $50.60\pm16.20$  \\
  SCAD     & $\phantom{0}74.91\pm 13.56$       & $55.00\pm30.98$    & $173.19\pm 50.32$       & $56.20\pm15.35$  \\ 
    MCP     & $\phantom{0}77.36\pm 13.42$       & $54.80\pm38.15$    & $177.97\pm 51.46$       & $49.40\pm\phantom{0}6.65$  \\ 
    \fos    & $\phantom{00}0.21\pm \phantom{0}0.09$       & $\phantom{0}1.00\pm\phantom{0}3.16$ & $\phantom{00}0.53\pm \phantom{0}0.14$       & $\phantom{0}1.30\pm\phantom{0}4.11$\\
        \toprule
        & \multicolumn{4}{c}{$~\nobs=500,\,p=500,s=10,\rho=0.3$}\\
    & \multicolumn{2}{c}{$M=2000$} & \multicolumn{2}{c}{$M=4000$}     \\
    \cline{2-5}
   Method     & Timing     & Hamming distance & Timing     & Hamming distance\\
    
     \lcvs\     & NA      & NA    & NA      & NA   \\
     SCAD     & $324.21\pm 107.43$       & $67.00\pm13.29$    & NA      & NA   \\ 
    MCP     & $361.54\pm 121.90$       & $57.50\pm\phantom{0}5.74$   & NA      & NA   \\ 
    \fos     & $\phantom{00}1.91\pm \phantom{00}0.52$      & $\phantom{0}1.30\pm\phantom{0}4.11$ & $\phantom{0}13.88\pm \phantom{0}3.85$       & $\phantom{0}1.50\pm\phantom{0}4.74$  \\
        \bottomrule
    
  \end{tabular}

\end{table*}

 \begin{table*}
\centering
 \caption{the average run times (in seconds) and average Hamming distances in these settings illustrate that \logfos\ performs well across a wide spectrum of correlations }
\label{tab:time_hamming-log-adddata}

  \begin{tabular}{c c c c c}
 
    \toprule 
      & \multicolumn{4}{c}{$~\nobs=200,\,p=500,s=15$} \\ 
    & \multicolumn{2}{c}{$\rho=0.0$} & \multicolumn{2}{c}{$\rho=0.25$}                  \\
	\cline{2-5}
    Method     & Timing     & Hamming distance & Timing     & Hamming distance\\
     \loglcvs\     & $19.97\pm \phantom{0}0.76$       & $11.50\pm\phantom{0}5.44$    & $22.94\pm \phantom{0}0.70$       & $14.60\pm\phantom{0}3.94$  \\

    \logfos    & $\phantom{0}0.89\pm \phantom{0}0.10$       & $\phantom{0}9.30\pm\phantom{0}4.19$ & $\phantom{0}0.86\pm \phantom{0}0.05$       & $10.60\pm\phantom{0}6.22$\\
        \toprule
        & \multicolumn{4}{c}{$~\nobs=200,\,p=500,s=15$}\\
    & \multicolumn{2}{c}{$\rho=0.5$} & \multicolumn{2}{c}{$\rho=0.75$}  \\
    \cline{2-5}
    Method     & Timing     & Hamming distance & Timing     & Hamming distance\\
    
     \loglcvs\     & $24.61\pm \phantom{0}2.23$      & $15.40\pm\phantom{0}2.11$    & $24.24\pm \phantom{0}3.10$      & $16.70\pm\phantom{0}2.94$  \\

    \logfos     & $\phantom{0}0.85\pm \phantom{0}0.07$      & $10.90\pm\phantom{0}2.76$   & $\phantom{0}0.85\pm \phantom{0}0.07$      & $14.70\pm\phantom{0}2.35$ \\
     \toprule
    & \multicolumn{4}{c}{$~\nobs=1000,\,p=5000,s=15$}\\
    & \multicolumn{2}{c}{$\rho=0.0$} & \multicolumn{2}{c}{$\rho=0.25$}                  \\
	\cline{2-5}
    Method     & Timing     & Hamming distance & Timing     & Hamming distance\\
     \loglcvs\     & NA & NA    & NA & NA   \\

    \logfos     & $25.95\pm \phantom{0}3.93$       & $\phantom{0}7.20\pm\phantom{0}3.70$ & $30.50\pm \phantom{0}2.99$      & $\phantom{0}6.80\pm\phantom{0}2.14$ \\
    
     \toprule
    & \multicolumn{4}{c}{$~\nobs=1000,\,p=5000,s=15$}\\
    & \multicolumn{2}{c}{$\rho=0.5$} & \multicolumn{2}{c}{$\rho=0.75$}                  \\
	\cline{2-5}
    Method     & Timing     & Hamming distance & Timing     & Hamming distance\\
     \loglcvs\     & NA & NA    & NA & NA   \\

    \logfos     & $41.84\pm \phantom{0}4.67$       & $\phantom{0}8.30\pm\phantom{0}1.94$ & $33.89\pm \phantom{0}1.64$      & $12.40\pm\phantom{0}2.54$ \\
    \bottomrule
    
  \end{tabular}
 
\end{table*}

\begin{table*}
  \caption{Average run times (in seconds), Hamming distances, and Estimation errors for \CHICHI\ and  \fos\ with $M=100$.
The results illustrate that \fos\ is both faster and more accurate than~\CHICHI}
 \label{tab:Liner_CHI} 
\centering
  \begin{tabular}{ c c  c c c  c c}
    \toprule
  & \multicolumn{2}{c}{$~\nobs=500,\,p=1000$} & \multicolumn{4}{c}{$\nobs=5000,\,p=10\,000$}                  \\
	\cline{2-6}
   Method     & Timing     & Hamming distance & Estimation error & Timing      & Hamming distance & Estimation error\\
	\hline
    \CHICHI & $\phantom{}0.35\pm \phantom{}0.13$  &  $2.30\pm0.82$  & $0.39\pm0.02$ & $63.28\pm11.20$  &$0.60\pm0.84$ & $0.31\pm0.02$\\
    \fos\     & $\phantom{}0.08\pm \phantom{}0.07$       & $\phantom{}1.00\pm\phantom{}2.82$ & $0.19\pm0.04$ & $\phantom{0}5.80\pm \phantom{0}3.06$      & $\phantom{}0.00\pm\phantom{}0.00$ & $0.10\pm0.02$ \\
    \bottomrule
  \end{tabular}
\end{table*}

 \begin{table*}
\centering
\caption{Average run times (in seconds), Hamming distances, and Estimation errors  for \LI\  and  \logfos\ with $M=500$.  
The results illustrate that \logfos\ is  faster and similarly accurate} 
\label{tab:log-reg-LI}
  \begin{tabular}{c c  c c c c c}
 
    \toprule 
      & \multicolumn{5}{c}{$~\nobs=200,\,p=200,s=8$} \\ 
    & \multicolumn{2}{c}{$\rho=0.25$} & \multicolumn{4}{c}{$\rho=0.5$}                  \\
	\cline{2-7}
    Method     & Timing     & Hamming distance & Estimation error & Timing     & Hamming distance & Estimation error \\
     \LI\     & $2.09\pm 0.33$       & $2.90\pm1.19$ &$2.25\pm0.60$    & $2.80\pm 0.37$       & $5.3\pm1.63$ &$2.50\pm0.68$   \\

    \logfos    & $\phantom{}0.36\pm0.06$       & $3.10\pm1.79$  &$2.04\pm0.59$  & $\phantom{}0.58\pm 0.03$       & $\phantom{}5.6\pm2.11$  &$2.35\pm0.69$ \\
        \toprule
        & \multicolumn{5}{c}{$~\nobs=200,\,p=500,s=8$}\\
    & \multicolumn{2}{c}{$\rho=0.25$} & \multicolumn{4}{c}{$\rho=0.5$}  \\
    \cline{2-7}
    Method     & Timing     & Hamming distance & Estimation error & Timing     & Hamming distance  & Estimation error \\
    
     \LI\     & $3.89\pm 0.46$      & $3.6\pm1.34$  &$2.57\pm0.17$   & $6.43\pm 0.89$      & $5.5\pm1.58$  &$2.73\pm0.34$\\

    \logfos     & $\phantom{}1.22\pm0.07$      & $\phantom{}4.4\pm1.95$  &$2.36\pm 0.16$  & $\phantom{}1.09\pm 0.04$      & $\phantom{}8.6\pm3.83$ &$2.56\pm0.36$ \\
     \toprule
    & \multicolumn{5}{c}{$~\nobs=1000,\,p=5000,s=8$}\\
    & \multicolumn{2}{c}{$\rho=0.25$} & \multicolumn{4}{c}{$\rho=0.5$}                  \\
	\cline{2-7}
    Method     & Timing     & Hamming distance & Estimation error  & Timing     & Hamming distance & Estimation error\\
     \LI\     & $142.50\pm53.98$ & $7.9\pm 0.31$  & $2.93\pm0.58$  & $314.29\pm130.60$ & $7.0\pm1.05$  & $2.98\pm0.58$ \\

    \logfos     & $\phantom{0}33.64\pm\phantom{0}4.69$       & $\phantom{}4.7\pm1.33$  & $2.48\pm0.58$ & $\phantom{0}25.95\pm \phantom{00}4.23$      & $\phantom{}3.8\pm1.68$  & $2.54\pm 0.57$ \\
    \bottomrule
    
  \end{tabular}

\end{table*}

\begin{table*}
\caption{Average run times (in seconds), Hamming distances, and  estimation errors  for~\fos, with $M=100$.
The results illustrate the  limited influence of the specific choice of~$\re$}
 \label{Const_impact_LR} 
\centering
  \begin{tabular}{ c c  c  c c}
    \toprule
  & \multicolumn{2}{c}{$~\nobs=500,\,p=1000,~\compatibility=2$}  \\
	\cline{2-4}
   Method     & Timing     & Hamming distance & Estimation error\\
	\hline
	 \fos ($\re=1$)\     & $0.12\pm 0.08$       & $1.0\pm2.82$ & $0.19\pm0.04$ \\
    \fos ($\re=2$)\     & $0.09\pm 0.05$       & $0.6\pm1.57$ & $0.19\pm0.04$ \\
     \fos ($\re=4$)\      & $0.09\pm 0.02$       & $0.6\pm1.57$ & $0.20\pm0.04$ \\    \bottomrule
  \end{tabular}
\end{table*}
\begin{table*}
\caption{Average run times (in seconds), Hamming distances, and  estimation errors  for~\fos, with $M=100$.
The results illustrate the  limited influence of the specific choice of~$\compatibility$}
 \label{C_impact_LR} 
\centering
  \begin{tabular}{ c c  c  c c}
    \toprule
  & \multicolumn{2}{c}{$~\nobs=500,\,p=1000,~\re=1$}  \\
	\cline{2-4}
   Method     & Timing     & Hamming distance & Estimation error\\
	\hline
	 \fos ($\compatibility=2$)\     & $0.12\pm 0.08$       & $1.0\pm2.82$ & $0.19\pm0.04$ \\
    \fos ($\compatibility=3$)\     & $0.04\pm 0.01$       & $0.3\pm0.67$ & $0.31\pm0.03$ \\
     \fos ($\compatibility=4$)\      & $0.03\pm 0.01$       & $2.4\pm1.26$ & $0.42\pm0.04$ \\    \bottomrule
  \end{tabular}
\end{table*}

\begin{table*}
\caption{Average run times (in seconds), Hamming distances, and  estimation errors for \logfos, with $M=500$.
The results illustrate the  limited influence of the specific choice of~$\re$}
 \label{Const_impact_Log_r} 
\centering
  \begin{tabular}{ c c  c  c c}
    \toprule
  & \multicolumn{2}{c}{$~\nobs=200,\,p=200,~\rho=0.25,~\constantlogreg=6$}  \\
	\cline{2-4}
   Method     & Timing     & Hamming distance & Estimation error\\
	\hline
	 \logfos ($\re=1$)\     & $0.36\pm \phantom{0}0.06$       & $3.1\pm1.79$ & $2.04\pm0.59$ \\
    \logfos ($\re=2$)\     & $0.78\pm \phantom{0}0.06$       & $4.4\pm1.95$ & $2.36\pm0.16$ \\
     \logfos ($\re=4$)\      & $0.73\pm \phantom{0}0.02$       & $4.4\pm1.95$ & $2.36\pm0.16$ \\    \bottomrule
  \end{tabular}
\end{table*}
\begin{table*}
\caption{Average run times (in seconds), Hamming distances, and  estimation errors for \logfos, with $M=500$.
The results illustrate the  limited influence of the specific choice of~$\constantlogreg$}
 \label{C_impact_Log_r} 
\centering
  \begin{tabular}{ c c  c  c c}
    \toprule
  & \multicolumn{2}{c}{$~\nobs=200,\,p=200,~\rho=0.25,~\re=1$}  \\
	\cline{2-4}
   Method     & Timing     & Hamming distance & Estimation error\\
	\hline
	 \logfos ($\constantlogreg=6$)\     & $0.36\pm \phantom{}0.06$       & $3.1\pm1.79$ & $2.04\pm0.59$ \\
    \logfos ($\constantlogreg=8$)\     & $0.33\pm \phantom{}0.05$       & $4.0\pm1.76$ & $1.95\pm0.59$ \\
     \logfos ($\constantlogreg=4$)\      & $0.32\pm \phantom{}0.02$       & $2.9\pm0.99$ & $2.17\pm0.60$ \\    \bottomrule
  \end{tabular}
\end{table*}

\section*{Acknowledgment}
We thank Haim Bar for the insightful comments on a draft version of this manuscript.
We also thank the associate editor and the reviewers for their insightful comments.
The authors MT and JL acknowledge funding from the Deutsche Forschungsgemeinschaft (DFG) under project number 265592081.

\bibliographystyle{IEEEtranN}
\bibliography{References}

\begin{thebibliography}{42}
\providecommand{\natexlab}[1]{#1}
\providecommand{\url}[1]{#1}
\csname url@samestyle\endcsname
\providecommand{\newblock}{\relax}
\providecommand{\bibinfo}[2]{#2}
\providecommand{\BIBentrySTDinterwordspacing}{\spaceskip=0pt\relax}
\providecommand{\BIBentryALTinterwordstretchfactor}{4}
\providecommand{\BIBentryALTinterwordspacing}{\spaceskip=\fontdimen2\font plus
\BIBentryALTinterwordstretchfactor\fontdimen3\font minus
  \fontdimen4\font\relax}
\providecommand{\BIBforeignlanguage}[2]{{%
\expandafter\ifx\csname l@#1\endcsname\relax
\typeout{** WARNING: IEEEtranN.bst: No hyphenation pattern has been}%
\typeout{** loaded for the language `#1'. Using the pattern for}%
\typeout{** the default language instead.}%
\else
\language=\csname l@#1\endcsname
\fi
#2}}
\providecommand{\BIBdecl}{\relax}
\BIBdecl

\bibitem[Tibshirani(1996)]{lasso}
R.~Tibshirani, ``Regression shrinkage and selection via the lasso,'' \emph{J.\@
  R.\@ Stat.\@ Soc.\@ Ser.\@ B Stat.\@ Methodol.}, vol.~58, no.~1, pp.
  267--288, 1996.

\bibitem[Bakin(1999)]{bakin1999}
S.~Bakin, ``Adaptive regression and model selection in data mining problems,''
  1999.

\bibitem[Yuan and Lin(2006)]{YuanLin06}
M.~Yuan and Y.~Lin, ``Model selection and estimation in regression with grouped
  variables,'' \emph{J.\@ Roy.\@ Statist.\@ Soc.\@ Ser.\@ B.}, vol.~68, no.~1,
  pp. 49--67, 2006.

\bibitem[Lederer(2022)]{LedererBook}
J.~Lederer, \emph{Fundamentals of high-dimensional statistics---with exercises
  and {R} labs}.\hskip 1em plus 0.5em minus 0.4em\relax Springer Series in
  Statistics, 2022.

\bibitem[B{\"u}hlmann and van~de Geer(2011)]{Buhlmann11}
P.~B{\"u}hlmann and S.~van~de Geer, \emph{Statistics for high-dimensional data:
  methods, theory and applications}, ser. Springer Ser.\@ Statist., 2011.

\bibitem[Wainwright(2019)]{wainwright2019high}
M.~Wainwright, \emph{High-dimensional statistics: a non-asymptotic
  viewpoint}.\hskip 1em plus 0.5em minus 0.4em\relax Cambridge Univ.\@ Press,
  2019, vol.~48.

\bibitem[Chichignoud et~al.(2016)Chichignoud, Lederer, and
  Wainwright]{Chichignoud_Lederer_Wainwright14}
M.~Chichignoud, J.~Lederer, and M.~Wainwright, ``A practical scheme and fast
  algorithm to tune the lasso with optimality guarantees,'' \emph{J.\@ Mach.\@
  Learn.\@ Res.}, vol.~17, no.~1, pp. 8162--8188, 2016.

\bibitem[Ch{\'e}telat et~al.(2017)Ch{\'e}telat, Lederer, and
  Salmon]{chetelat2017optimal}
D.~Ch{\'e}telat, J.~Lederer, and J.~Salmon, ``Optimal two-step prediction in
  regression,'' \emph{Electron.\@ J.\@ Stat.}, vol.~11, no.~1, pp. 2519--2546,
  2017.

\bibitem[Laszkiewicz et~al.(2021)Laszkiewicz, Fischer, and
  Lederer]{laszkiewicz2021thresholded}
M.~Laszkiewicz, A.~Fischer, and J.~Lederer, ``Thresholded adaptive validation:
  Tuning the graphical lasso for graph recovery,'' in \emph{International
  Conference on Artificial Intelligence and Statistics}.\hskip 1em plus 0.5em
  minus 0.4em\relax PMLR, 2021, pp. 1864--1872.

\bibitem[Huang et~al.(2021)Huang, D{\"u}ren, Hellton, and
  Lederer]{huang2021tuning}
S.~Huang, Y.~D{\"u}ren, K.~Hellton, and J.~Lederer, ``Tuning parameter
  calibration for personalized prediction in medicine,'' \emph{Electron.\@ J.\@
  Stat.}, vol.~15, no.~2, pp. 5310--5332, 2021.

\bibitem[Li and Lederer(2019)]{Li2019}
W.~Li and J.~Lederer, ``Tuning parameter calibration for $\ell_{1}$-regularized
  logistic regression,'' \emph{J.\@ Statist.\@ Plann.\@ Inference}, vol. 202,
  pp. 80--98, 2019.

\bibitem[Judson et~al.(2002)Judson, Salisbury, Schneider, Windemuth, and
  Stephens]{judson2002many}
R.~Judson, B.~Salisbury, J.~Schneider, A.~Windemuth, and J.~Stephens, ``How
  many {SNP}s does a genome-wide haplotype map require?''
  \emph{Pharmacogenomics}, vol.~3, no.~3, pp. 379--391, 2002.

\bibitem[Bunea et~al.(2013)Bunea, Lederer, and She]{Bunea}
F.~Bunea, J.~Lederer, and Y.~She, ``The group square-root lasso: Theoretical
  properties and fast algorithms,'' \emph{IEEE Trans.\@ Inform.\@ Theory},
  vol.~60, no.~2, pp. 1313--1325, 2013.

\bibitem[Friedman et~al.(2010)Friedman, Hastie, and Tibshirani]{Hastie10}
J.~Friedman, T.~Hastie, and R.~Tibshirani, ``Regularization paths for
  generalized linear models via coordinate descent,'' \emph{J.\@ Stat.\@
  Softw.}, vol.~33, no.~1, pp. 1--22, 2010.

\bibitem[Osborne et~al.(2000)Osborne, Presnell, and Turlach]{osborne2000lasso}
M.~Osborne, B.~Presnell, and B.~Turlach, ``On the lasso and its dual,''
  \emph{J.\@ Comput.\@ Graph.\@ Statist.}, vol.~9, no.~2, pp. 319--337, 2000.

\bibitem[Hu et~al.(2017)Hu, Li, Meng, Qin, and Yang]{hu2017group}
Y.~Hu, C.~Li, K.~Meng, J.~Qin, and X.~Yang, ``Group sparse optimization via
  $\ell_{p, q}$ regularization,'' \emph{J.\@ Mach.\@ Learn. \@ Res.}, vol.~18,
  no.~1, pp. 960--1011, 2017.

\bibitem[Negahban et~al.(2012)Negahban, Ravikumar, Wainwright, and
  Yu]{Negahban12}
S.~Negahban, P.~Ravikumar, M.~Wainwright, and B.~Yu, ``A unified framework for
  high-dimensional analysis of ${M}$-estimators with decomposable
  regularizers,'' \emph{Statist.\@ Sci.}, vol.~27, no.~4, pp. 538--557, 2012.

\bibitem[Wainwright(2014)]{Wainwright14}
M.~Wainwright, ``Structured regularizers for high-dimensional problems:
  statistical and computational issues,'' \emph{Annu.\@ Rev.\@ Stat.\@ Appl.},
  vol.~1, pp. 233--253, 2014.

\bibitem[Koltchinskii(2009)]{koltch09a}
V.~Koltchinskii, ``Sparsity in penalized empirical risk minimization,''
  \emph{Ann.\@ Inst.\@ Henri Poincar\'e Probab.\@ Stat.}, vol.~45, no.~1, pp.
  7--57, 2009.

\bibitem[van~de Geer and B{\"u}hlmann(2009)]{Sara09}
S.~van~de Geer and P.~B{\"u}hlmann, ``On the conditions used to prove oracle
  results for the lasso,'' \emph{Electron.\@ J.\@ Stat.}, vol.~3, pp.
  1360--1392, 2009.

\bibitem[Lederer and Vogt(2021)]{lederer2021estimating}
J.~Lederer and M.~Vogt, ``Estimating the lasso’s effective noise,''
  \emph{JMLR}, vol.~22, no. 276, pp. 1--32, 2021.

\bibitem[Lederer et~al.(2019)Lederer, Yu, and Gaynanova]{Lederer2019}
J.~Lederer, L.~Yu, and I.~Gaynanova, ``Oracle inequalities for high-dimensional
  prediction,'' \emph{Bernoulli}, vol.~25, no.~2, pp. 1225--1255, 2019.

\bibitem[Dalalyan et~al.(2017)Dalalyan, Hebiri, and
  Lederer]{dalalyan2017prediction}
A.~Dalalyan, M.~Hebiri, and J.~Lederer, ``On the prediction performance of the
  lasso,'' \emph{Bernoulli}, vol.~23, no.~1, pp. 552--581, 2017.

\bibitem[Hastie et~al.(2015)Hastie, Tibshirani, and Wainwright]{Hastie2015}
T.~Hastie, R.~Tibshirani, and M.~Wainwright, \emph{Statistical learning with
  sparsity: the lasso and generalizations}.\hskip 1em plus 0.5em minus
  0.4em\relax CRC press, 2015.

\bibitem[Borwein and Lewis(2010)]{Borwein2010}
J.~Borwein and A.~Lewis, \emph{Convex analysis and nonlinear optimization:
  theory and examples}.\hskip 1em plus 0.5em minus 0.4em\relax Springer Science
  \& Business Media, 2010.

\bibitem[Beck and Teboulle(2009)]{Beck09}
A.~Beck and M.~Teboulle, ``A fast iterative shrinkage-thresholding algorithm
  for linear inverse problems,'' \emph{SIAM J.\@ Imaging Sci.}, vol.~2, no.~1,
  pp. 183--202, 2009.

\bibitem[Fan and Li(2001)]{Fan_Li01}
J.~Fan and R.~Li, ``Variable selection via nonconcave penalized likelihood and
  its oracle properties,'' \emph{J.\@ Amer.\@ Statist.\@ Assoc.}, vol.~96, no.
  456, pp. 1348--1360, 2001.

\bibitem[Zhang(2010)]{Zhang10}
C.~Zhang, ``Nearly unbiased variable selection under minimax concave penalty,''
  \emph{Ann.\@ Statist.}, vol.~38, no.~2, pp. 894--942, 2010.

\bibitem[Bach et~al.(2011{\natexlab{a}})Bach, Jenatton, Mairal, and
  Obozinski]{BJMO11}
F.~Bach, R.~Jenatton, J.~Mairal, and G.~Obozinski, ``Convex optimization with
  sparsity-inducing norms,'' in \emph{Optimization for Machine Learning}.\hskip
  1em plus 0.5em minus 0.4em\relax MIT Press, 2011.

\bibitem[Zhou et~al.(2012)Zhou, Armagan, and Dunson]{Zhou12}
H.~Zhou, A.~Armagan, and D.~Dunson, ``Path following and empirical bayes model
  selection for sparse regression,'' \emph{arXiv:1201.3528.}, 2012.

\bibitem[Kogan et~al.(2009)Kogan, Levin, Routledge, Sagi, and Smith]{Kogan09}
S.~Kogan, D.~Levin, B.~Routledge, J.~Sagi, and N.~Smith, ``Predicting risk from
  financial reports with regression,'' in \emph{HLT-NAACL}, 2009, pp. 272--280.

\bibitem[Guyon et~al.(2008)Guyon, Aliferis, Cooper, Elisseeff, Pellet, Spirtes,
  and Statnikov]{Guyon08}
I.~Guyon, C.~Aliferis, G.~Cooper, A.~Elisseeff, J.~Pellet, P.~Spirtes, and
  A.~Statnikov, ``Design and analysis of the causation and prediction
  challenge,'' in \emph{WCCI Causation and Prediction Challenge}, 2008, pp.
  1--33.

\bibitem[Meinshausen and B{\"u}hlmann(2006)]{Meinshausen06}
N.~Meinshausen and P.~B{\"u}hlmann, ``High-dimensional graphs and variable
  selection with the lasso,'' \emph{Ann.\@ Statist.}, vol.~34, no.~3, pp.
  1436--1462, 2006.

\bibitem[Statnikov et~al.(2015)Statnikov, Ma, Henaff, Lytkin, Efstathiadis,
  Peskin, and Aliferis]{Statnikov15}
A.~Statnikov, S.~Ma, M.~Henaff, N.~Lytkin, E.~Efstathiadis, E.~Peskin, and
  C.~Aliferis, ``Ultra-scalable and efficient methods for hybrid observational
  and experimental local causal pathway discovery,'' \emph{J.\@ Mach.\@
  Learn.\@ Res.}, vol.~16, no.~1, pp. 3219--3267, 2015.

\bibitem[Ma et~al.(2004)Ma, Wang, Ryan, Isakoff, Barmettler, Fuller, Muir,
  Mohapatra, Salunga, Tuggle, and Tran]{Xiao}
X.~Ma, Z.~Wang, P.~Ryan, S.~Isakoff, A.~Barmettler, A.~Fuller, B.~Muir,
  G.~Mohapatra, R.~Salunga, J.~Tuggle, and Y.~Tran, ``A two-gene expression
  ratio predicts clinical outcome in breast cancer patients treated with
  tamoxifen,'' \emph{Cancer cell}, vol.~5, no.~6, pp. 607--616, 2004.

\bibitem[Subramanian et~al.(2005)Subramanian, Tamayo, Mootha, Mukherjee, Ebert,
  Gillette, Paulovich, Pomeroy, Golub, Lander, and Mesirov]{GSEA}
A.~Subramanian, P.~Tamayo, V.~Mootha, S.~Mukherjee, B.~Ebert, M.~Gillette,
  A.~Paulovich, S.~Pomeroy, T.~Golub, E.~Lander, and J.~Mesirov, ``Gene set
  enrichment analysis: a knowledge-based approach for interpreting genome-wide
  expression profiles,'' \emph{Proc.\@ Nat.\@ Acad.\@ Sci.\@ India Sect.\@ A},
  vol. 102, no.~43, pp. 15\,545--15\,550, 2005.

\bibitem[Zhao and Yu(2006)]{Zhao06}
P.~Zhao and B.~Yu, ``On model selection consistency of lasso,'' \emph{J.\@
  Mach.\@ Learn.\@ Res.}, vol.~7, pp. 2541--2563, 2006.

\bibitem[Mian et~al.(2005)Mian, Ugurel, Parkinson, Schlenzka, Dryden,
  Lancashire, Ball, Creaser, Rees, and Schadendorf]{mian2005serum}
S.~Mian, S.~Ugurel, E.~Parkinson, I.~Schlenzka, I.~Dryden, L.~Lancashire,
  G.~Ball, C.~Creaser, R.~Rees, and D.~Schadendorf, ``Serum proteomic
  fingerprinting discriminates between clinical stages and predicts disease
  progression in melanoma patients,'' \emph{J.\@ Clin.\@ Oncol.}, vol.~23,
  no.~22, pp. 5088--5093, 2005.

\bibitem[Lederer and M{\"u}ller(2015)]{lederer2015don}
J.~Lederer and C.~M{\"u}ller, ``Don't fall for tuning parameters: tuning-free
  variable selection in high dimensions with the {TREX},'' in
  \emph{Twenty-Ninth AAAI Conference on Artificial Intelligence}, 2015.

\bibitem[Bach et~al.(2011{\natexlab{b}})Bach, Jenatton, Mairal, and
  Obozinski]{bach2011optimization}
F.~Bach, R.~Jenatton, J.~Mairal, and G.~Obozinski, ``Optimization with
  sparsity-inducing penalties,'' \emph{arXiv:1108.0775.}, 2011.

\bibitem[Ndiaye et~al.(2016)Ndiaye, Fercoq, Gramfort, and Salmon]{Ndiaye16}
E.~Ndiaye, O.~Fercoq, A.~Gramfort, and J.~Salmon, ``Gap safe screening rules
  for sparse-group lasso,'' in \emph{NIPS}, 2016, pp. 388--396.

\bibitem[Ghaoui et~al.(2010)Ghaoui, Viallon, and Rabbani]{Ghaoui2010}
L.~Ghaoui, V.~Viallon, and T.~Rabbani, ``Safe feature elimination for the lasso
  and sparse supervised learning problems,'' \emph{arXiv:1009.4219.}, 2010.

\end{thebibliography}
\begin{IEEEbiographynophoto}{Mahsa Taheri} received her Master’s degree in Artificial Intelligence from IUT, Iran, in 2017. 
She is currently pursuing a Ph.D. degree in Mathematical Statistics at the Ruhr-University Bochum.
Her research interests include statistics and theoretical machine learning, specifically,  theory of deep learning. 
\end{IEEEbiographynophoto}
\begin{IEEEbiographynophoto}{Néhémy Lim} is a Data Engineering consultant at Business and Decision. His research interests include high-dimensional statistics, time series forecasting, and machine learning with applications in biology. He received the Ph.D. degree in Computer Science from Université d'Évry-Val-d'Essonne in 2015. He was a postdoctoral research associate at the University of Washington from 2015 to 2017 and a visiting assistant professor at the University of Connecticut from 2017 to 2020.
\end{IEEEbiographynophoto}
\begin{IEEEbiographynophoto}{Johannes Lederer}
 is Professor of Mathematical Statistics at the Ruhr-University Bochum. His educational training spans physics, mathematics, and statistics. After receiving a MSc in Physics, he started his Doctor of Sciences under the supervision of Sara van de Geer and Peter Bühlmann at ETH Zürich in December 2009 and completed the degree in December 2012. After that, he was a post-doctoral researcher hosted by Martin Wainwright at UC Berkeley, until he accepted the position as the Jacob Wolfowitz Visiting Assistant Professor in the Department of Statistical Science at Cornell University in August 2013. In September 2015, he joined the University of Washington as Tenure-Track Assistant Professor of Statistics and Adjunct Assistant Professor of Biostatistics at the University of Washington. In December 2017, Johannes was appointed in Bochum. Johannes is also Affiliate Faculty at the University of Washington and Associate Editor of the Canadian Journal of Statistics.\end{IEEEbiographynophoto}

\vskip 0.2in

\end{document}